\documentclass[a4paper]{article}

\usepackage{bbm}
\usepackage{latexsym}
\usepackage{amssymb}
\usepackage{graphicx}
\usepackage{psfrag}
\usepackage{verbatim}
\usepackage[mathcal]{euscript}

\newcommand{\be}{\begin{equation}}
\newcommand{\ee}{\end{equation}}
\newcommand{\bea}{\begin{eqnarray}}
\newcommand{\eea}{\end{eqnarray}}
\newcommand{\bean}{\begin{eqnarray*}}
\newcommand{\eean}{\end{eqnarray*}}

\renewcommand{\b}{\langle}
\renewcommand{\k}{\rangle}

\newcommand{\irm}{{\rm i}}

\renewcommand{\d}{{\rm d}}
\newcommand{\cl}[1]{{\cal #1}}

\newcommand{\pa}{\partial}

\newcommand{\ts}{\textstyle}
\newcommand{\sst}{\scriptstyle}
\newcommand{\ssst}{\scriptscriptstyle}

\newcommand{\pdiff}[2]{\frac{\partial #1}{\partial #2}}

\newcommand{\bN}{\mathbbm{N}}
\newcommand{\bC}{\mathbbm{C}}
\newcommand{\bR}{\mathbbm{R}}

\newcommand{\clH}{\cl{H}}
\newcommand{\clL}{\cl{L}}

\newcommand{\clS}{\cl{S}}
\newcommand{\clF}{F}

\newcommand{\clA}{\cl{A}}
\newcommand{\clR}{\cl{R}}

\newcommand{\eq}[1]{(\ref{#1})}

\newcommand{\fig}[1]{Fig.\ \ref{#1}}

\newcommand{\pic}[4]
{
 \begin{figure}
 \begin{center}
 \includegraphics[height=#3]{#4}
 \end{center}
 \caption{\label{#1} #2}
 \end{figure}
}

\newcommand{\tr}{{\rm tr}}

\newcommand{\FBC}{\clF^{\rm\ssst BC}}

\newcommand{\Phit}{\tilde{\Phi}}
\newcommand{\Omegat}{\tilde{\Omega}}
\newcommand{\Ft}{\tilde{\clF}}
\newcommand{\St}{\tilde S}
\newcommand{\At}{\tilde{A}}
\newcommand{\FBCt}{\Ft^{\rm\ssst BC}}
\newcommand{\Gammat}{\tilde{\Gamma}}

\begin{document}
\thispagestyle{empty}
\hfill
\parbox[t]{3.6cm}{
hep-th/0504059 \\
AEI-2005-090} \vspace{2cm}

\begin{center}
{\bf\Large Geometric spin foams, Yang-Mills theory and background-independent models} \\[4mm]
{Florian Conrady \\[2mm]
\small\it Max-Planck-Institut f\"{u}r Gravitationsphysik, Albert-Einstein-Institut, \\ D-14476 Golm, Germany \\[1mm]
\small\it Centre de Physique Th\'{e}orique de Luminy, CNRS, \\ F-13288 Marseille,
France}
\end{center}
\vskip.5cm

\begin{abstract}
\noindent
We review the dual transformation from pure lattice gauge theory to spin foam models with an emphasis on a geometric viewpoint. This allows us to give a simple dual formulation of SU(N) Yang-Mills theory, where spin foam surfaces are weighted with the exponentiated area. In the case of gravity, we introduce a symmetry condition which demands that the amplitude of an individual spin foam depends only on its geometric properties and not on the lattice on which it is defined. For models that have this property, we define a new sum over abstract spin foams that is independent of any choice of lattice or triangulation. We show that a version of the Barrett-Crane model satisfies our symmetry requirement.
\end{abstract}
\vskip1cm

\section{Introduction}
\label{introduction}

The concept of spin foams is both new and old. It is old in the sense that it is just another name for the plaquette diagrams that appear in the strong coupling expansion of lattice gauge theories---a method that has been developed in the seventies (see e.g.\ \cite{DrouffeZuber}). It is also new, however, because it was reinvented in the nineties within the context of quantum gravity, and used to construct rigorous proposals for sums over geometries \cite{Reisenbergerworldsheet,IwasakireformulationofPonzanoRegge}.

In both cases, the sum over spin foam diagrams arises from character expansions on plaquettes, and a subsequent integration over group variables. Although this is a well-known procedure in the strong coupling expansion, it is less known that it can be also used to give an alternative {\it non-perturbative} definition of the original theory.
For pure lattice gauge theory, this exact transformation from path integral to spin foam sum has been described by Oeckl \& Pfeiffer \cite{OecklPfeifferdualofpurenonAbelian}.

In the first part of the article, we explain this {\it dual} transformation in a pedagogic and step-by-step fashion. Our presentation differs from that in \cite{OecklPfeifferdualofpurenonAbelian} in that we lay emphasis on the fact that plaquettes can be organized into larger surfaces with a single representation label\footnote{This is known in lattice field theory \cite{DrouffeZuber} and has been pointed out in the spin foam literature \cite{ReisenbergerlatticefourdEuclidean,Baezspinfoammodels}.}. The different surfaces meet along lines, so the entire diagram takes the form of a branched surface with labels on its unbranched components. Thus, we are led to a more geometric definition of spin foams, where the spin foam is not identified with the lattice, but instead regarded as a branched surface that lies on it. Based on this geometric viewpoint, we obtain a particularly simple description of a spin foam model that is dual to lattice Yang-Mills theory in dimension $d\ge 2$: each unbranched component is weighted with $\exp(-T_\rho \clA)$, where $\clA$ is the area of the surface and $T_\rho$ a representation-dependent tension.

The second part of the paper concerns the spin foam approach to gravity: i.e.\ the attempt to employ spin foam sums for defining non-perturbative and background-independent models of quantum gravity. The construction of such models is usually plagued by the problem that it depends on the choice of a lattice or triangulation. The spin foam sum is restricted to spin foams that are congruent with the lattice. Unless the theory is topologically invariant, the latter stands in obvious contradiction to background and cutoff independence.

We propose a solution to this difficulty which is, again, based on the geometric notion of spin foams. There are spin foam models whose weight factors are independent of how spin foams are embedded on the lattice. Their amplitudes depend only on topological properties of the branched surface, but not on how its components and branching lines are subdivided by the lattice. We elevate this property to a symmetry principle, which appears natural from the relational point of view: what counts is how different surfaces connect to each other, while subdivisions within a surface are physically irrelevant. When a spin foam model satisfies the symmetry requirement, we discard the lattice and extend the model to a sum over {\it all} branched and labelled surfaces in the manifold. The new sum is lattice-independent, but contains an infinite overcounting of homeomorphically equivalent configurations. We factor this gauge volume off, and arrive at a sum over {\it abstract} spin foams---equivalence classes of spin foams under homeomorphisms---which carry only topological and combinatorial information. As an example, we consider a version of the Barrett-Crane model, for which we check the symmetry property. 

With this method, we propose an alternative to the group field theory approach \cite{ReisenbergerRovelliconnectionformulation, ReisenbergerRovellispinfoamsasFeynmandiagrams}, where background-independence is achieved by a sum over lattices. Related ideas on gravity spin foams have been expressed by Bojowald \& Perez \cite{BojowaldPerez} and Zapata \cite{Zapata}. Our abstract spin foams are similar to the combinatorial spin foams of the causal histories approach \cite{MaScausalevolution,MaSlocalcausality}. The dual Yang-Mills model should be compared with attempts to formulate lattice Yang-Mills theory in terms of strings (see e.g.\ \cite{Kostov}).

For readers who are not familiar with techniques of loop and spin foam gravity, we have included a short section on spin network states. Spin networks are a generalization of Wilson loops and provide a basis for functionals of the connection. Introductions to canonical loop quantum gravity and the spin foam approach can be found in \cite{Rovellibook,Thiemannintroduction} and \cite{Baezintroduction,Perezreview} respectively.

The paper is organized as follows: in the next section, we set our framework and conventions for pure lattice gauge theory. Section \ref{spinnetworkstates} gives the introduction to spin networks. The latter will be extensively used in section \ref{dualtransformationtothespinfoammodel} where we explain the dual transformation. It is also there that the geometric definition of spin foams and its motivation will become clear. After that we present the spin foam model of Yang-Mills theory (sec.\ \ref{aspinfoammodelofYangMillstheory}). In section \ref{backgroundindependentspinfoammodels}, we consider spin foam models of gravity: we introduce the symmetry condition and describe the extension to the lattice-independent formulation. We verify in the appendix that it is applicable to one of the versions of the Barrett-Crane model. Section \ref{summaryanddiscussion} contains the summary and discussion.

\section{Lattice gauge theory}
\label{latticegaugetheory}

Consider a hypercubic lattice $\kappa$ in $\bR^d$ that has lattice constant $a$ and finite side lengths $L_i$, $i=1,\ldots,n$. The dimension $d$ should be greater than 1. Let us choose an orientation for each link and plaquette of the lattice, and call the resulting oriented links and plaquettes {\it edges} $e$ and {\it faces} $f$ respectively. The choice of this orientation is arbitrary and all physical quantities are independent of it. 
We write $E_{\kappa}$ for the set of all edges of $\kappa$. 
The edges on the boundary of $\kappa$ form again a lattice which we denote by $\pa\kappa$.

On the lattice, connections are represented by functions
\be
g: E_{\kappa}\to G\,,\quad e\mapsto g_e
\ee
that map edges of $\kappa$ into elements of the gauge group $G$. We denote the configuration space of all connections on $\kappa$ by $\clA_\kappa$. The gauge group is assumed to be a compact Lie group.

The quantities of physical interest are path integrals
\be
W(\Phi) = \int\left({\ts\prod\limits_{e\in\kappa}}\d g_e\right) \exp\Big(\iota \clS(g)\Big)\Phi(g)\,.
\label{generalamplitude}
\ee
For each edge $e$, we integrate over $g_e$ by using the Haar measure on $G$.
The action $\clS$ is a real and gauge-invariant functional of the connection $g$, and $\Phi$ stands for a weighting functional. Depending on the context, the amplitude $W(\Phi)$ can be the partition function ($\Phi = 1$), the mean value of an observable $\Phi$ or a transition amplitude. $\iota$ is either $\irm$ or $-1$, depending on whether we define a Euclidean or Minkowskian version of the theory.

We keep the choice of $\iota$ unspecified, but require the action to have the form
\be
\clS = \sum_{f\in\kappa^\circ} \clS_f\,.
\ee
The sum runs over all faces in the interior $\kappa^\circ$ of the lattice. Each face action $\clS_f$ is gauge-invariant and depends only on group elements of edges surrounding the face $f$.

\psfrag{e1}{$e_1$}
\psfrag{e2}{$e_2$}
\psfrag{e3}{$e_3$}
\psfrag{e4}{$e_4$}
\psfrag{f}{$f$}
The standard example is Euclidean SU(N) Yang-Mills theory in $d$ dimensions, where the face action is chosen as the Wilson action
\be
\clS_f(g) = \frac{2 N}{a^{4-d}\gamma^2}\left[1 - \frac{1}{2N}\,\tr\left(U_f(g) + U^\dagger_f(g)\right)\right]\,.
\label{Wilsonaction}
\ee
Here, $U_f$ denotes the holonomy around the face. For face and edge orientations as in \fig{exampleofedgeorientationsaroundaplaquette}, it is given by
\be
U_f(g) = g^{-1}_{e_4}\,g^{-1}_{e_3}\,g_{e_2}\,g_{e_1}\,.
\ee
$\gamma$ is the gauge coupling and $\tr$ stands for the trace in the defining representation of SU(N).
The classical continuum limit of \eq{Wilsonaction} yields the Yang-Mills action
\be
\clS = \frac{1}{4\gamma^2}\int_M\d^d x\;F^a_{\mu\nu}F_a^{\mu\nu}\,.
\ee
In the quantum theory, one describes the limit $a\to 0$ by a sequence of lattice actions \eq{Wilsonaction}
that have an $a$-dependent gauge coupling $\gamma(a)$. 
\pic{exampleofedgeorientationsaroundaplaquette}{Example of edge orientations around a plaquette.}{2cm}{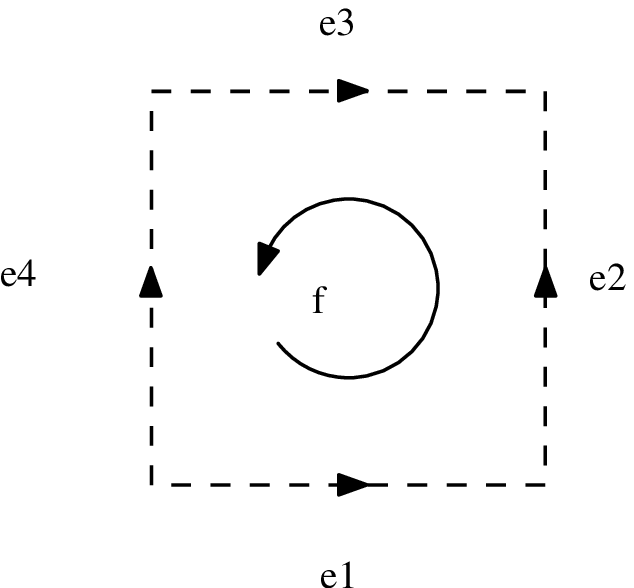}

Another important example is lattice BF theory where the exponentiated face action equals
\be
\exp\Big(\irm\,\clS_f(g)\Big) = \delta(U_f(g))\,.
\label{plaquetteactionBFtheory}
\ee
Its definition can be formally derived from a continuum path integral
\be
\int DA \int DB\; \exp\Big(\irm\,\tr\left[B F(A)\right]\Big) \sim \int DA\; \delta(F(A))\,.
\label{formalpathintegralofBFtheory}
\ee
where, in addition to the connection degrees of freedom, we have a Lie algebra valued two-form field $B$.

In this article, we are primarily interested in amplitudes \eq{generalamplitude} of boundary states. By boundary states we mean weighting functionals $\Phi$ that depend only on group elements of boundary edges. We write the associated amplitude as
\be
\Omega(\Phi) := \int\left({\ts\prod\limits_{e\in\kappa}}\d g_e\right)\;\exp\left(\iota \clS(g)\right)\Phi^*(\pa g)\,.
\label{boundaryamplitude}
\ee
$\pa g$ denotes the restriction of the connection $g$ to the boundary. We assume that $\Phi$ is an element in the Hilbert space $\clL^2_0(\clA_{\pa\kappa})$ of square-integrable and gauge-invariant boundary functionals, i.e.
\be
\int\left({\ts\prod\limits_{e\in\kappa}}\d g_e\right)\;\Phi^*(g)\Phi(g)\;<\; \infty\,.
\ee
Intuitively, one may think of $\Omega(\Phi)$ as the transition amplitude from ``nothing'' to the state $\Phi$. When $\Phi$ is a product $\Phi_1\Phi_2$ of functionals, where $\Phi_1$ and $\Phi_2$ have disjoint domains, we can think of \eq{boundaryamplitude} as the transition amplitude between $\Phi_1$ and $\Phi_2$. 

\section{Spin network states}
\label{spinnetworkstates}

A pure lattice gauge theory as described above can be transformed to a physically equivalent spin foam model. This so-called dual transformation rests on the use of a particular basis for functionals of connections: the basis of spin network states\footnote{The concept was first introduced in a context where the group was SU(2), hence the word ``spin''.}.

\psfrag{a}{(a)}
\psfrag{b}{(b)}
\psfrag{re1}{$\rho_{e_1}$}
\psfrag{re2}{$\rho_{e_2}$}
\psfrag{re3}{$\rho_{e_3}$}
\psfrag{Iv1}{$I_{v_1}$}
\psfrag{Iv2}{$I_{v_2}$}
\psfrag{r1out}{$\rho^{\ssst\rm out}_1$}
\psfrag{r2out}{$\rho^{\ssst\rm out}_2$}
\psfrag{r3out}{$\rho^{\ssst\rm out}_3$}
\psfrag{r1in}{$\rho^{\ssst\rm in}_1$}
\psfrag{r2in}{$\rho^{\ssst\rm in}_2$}
\psfrag{r3in}{$\rho^{\ssst\rm in}_3$}
\psfrag{Iv}{$I_v$}
Generically, the term {\it spin network} refers to a directed graph of valence $\ge 2$ that has a certain labelling (\fig{spinnetwork}a):  each edge $e$ is labelled by a unitary irreducible representation $\rho_e$ of the gauge group\footnote{More precisely, the $\rho_e$'s are taken from a set $\clR$ that contains one representative for each equivalence class of unitary irreducible representations of $G$.}, and each vertex $v$ carries an invariant tensor (or intertwiner) $I_v$ in the tensor product
\be
V_{\rho^{\ssst\rm out}_1}\otimes\cdots\otimes V_{\rho^{\ssst\rm out}_m}\otimes
V^*_{\rho^{\ssst\rm in}_1}\otimes\cdots\otimes V^*_{\rho^{\ssst\rm in}_n}\,.
\label{tensorproduct}
\ee
Here, each $V_\rho$ stands for the representation space of the irrep $\rho$.
For every outgoing edge $e^{\ssst\rm out}_i$ with label $\rho^{\ssst\rm out}_i$, the tensor space has a component $V_{\rho^{\ssst\rm out}_i}$, while for each incoming edge with label $\rho^{\ssst\rm in}_j$, we get the dual of the associated representation space (\fig{spinnetwork}b). We write $\rho = {\rm\sst triv}$ for the trivial representation. The representation labels are sometimes referred to as {\it colors} of edges.
\pic{spinnetwork}{Labelling of spin network graphs.}{3cm}{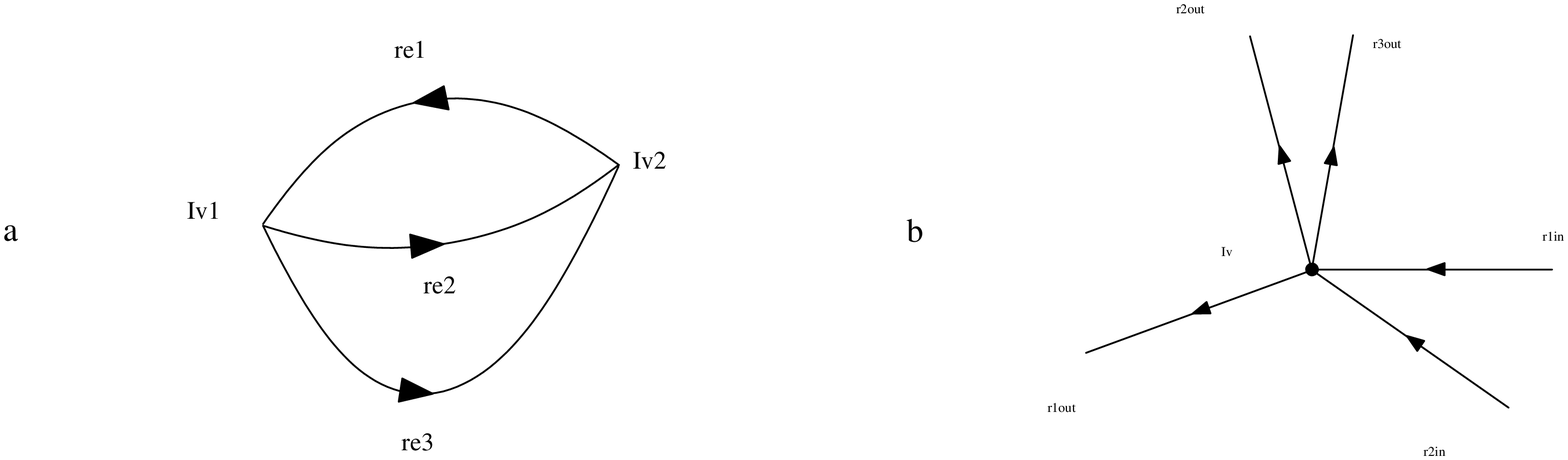}

To every spin network $S$ corresponds a {\it spin network functional} or {\it state}, which we denote by $\Psi_S$. For a given connection $g$, the value $\Psi_S(g)$ of the functional is obtained as follows:
for every edge $e$ of the spin network graph, there is a group element $g_e$ and its representation $\rho_e(g_e)$ in terms of the representation $\rho_e$. We contract all such representation maps with intertwiners, in the way indicated by the graph, and thereby receive a number---the value of $S$ on the connection $g$. For later convenience, we enhance this rule by adding a factor $(\dim V_\rho)^{1/2}$ for every edge of the spin network. In formulas, we can write the value $\Psi_S(g)$ as
\be
\Psi_S(g) = \left(\prod_{v\in\kappa} I_v\right){\bf\cdot}\left(\prod_{e\in\kappa}\left(\dim V_{\rho_e}\right)^{1/2} \rho_e(g_e)\right) 
\ee
where the dot $\bf\cdot$ symbolizes the contraction of tensor indices.

\psfrag{r1}{$\sst\rho_1$}
\psfrag{r2}{$\sst\rho_2$}
\psfrag{r3}{$\sst\rho_3$}
In the present case, the connection lives only on the lattice $\kappa$, so spin network graphs lie on $\kappa$
(\fig{spinnetworkonalattice}). When the orientation of the spin network edge is opposite to that of the lattice edge, the spin network functional receives a factor $\rho_e(g^{-1}_e)$ instead of $\rho_e(g_e)$.
\pic{spinnetworkonalattice}{Spin network on a lattice (intertwiner labels are omitted).}{5cm}{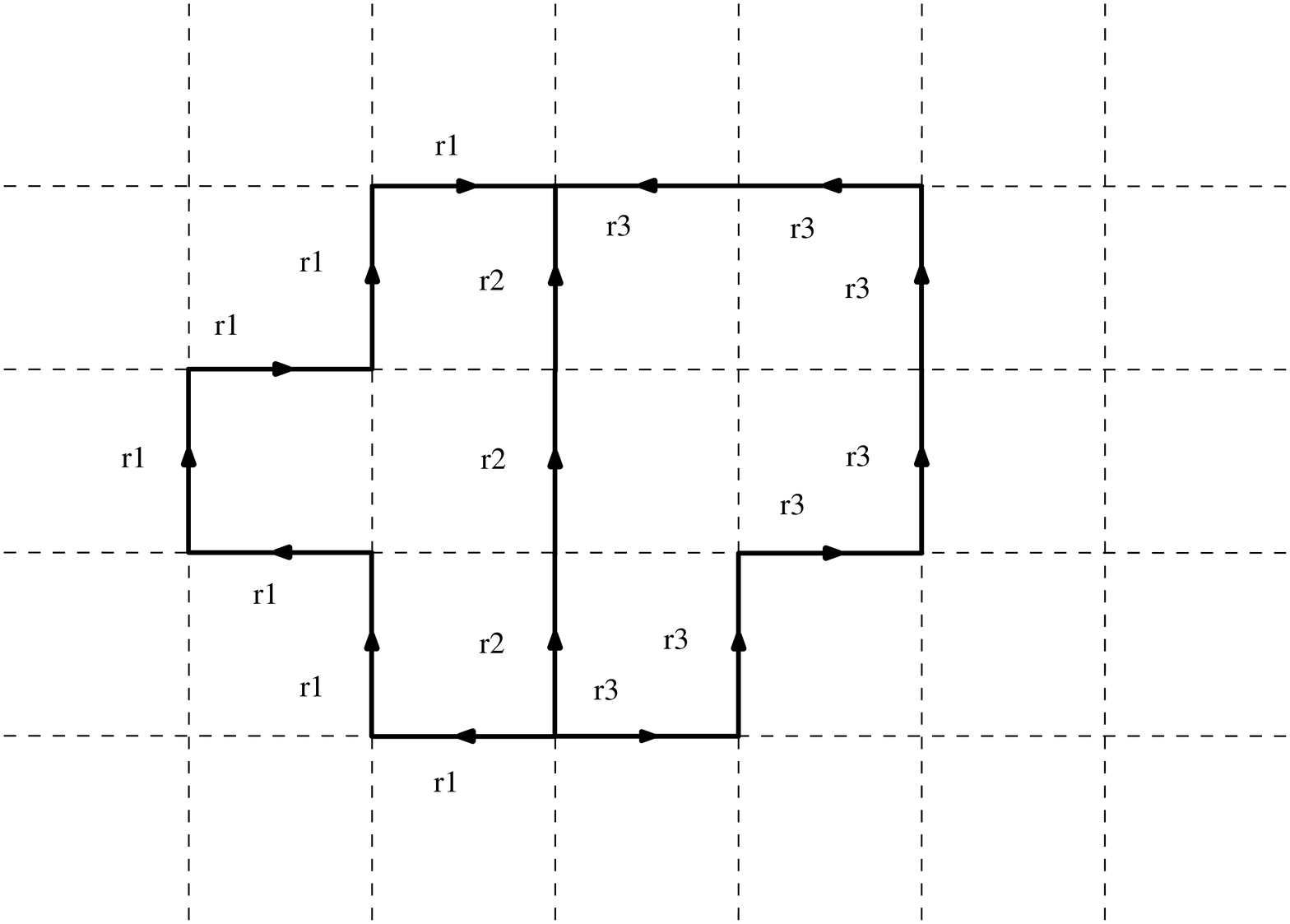}

\begin{figure}
\hspace{-2cm}\includegraphics[height=7cm]{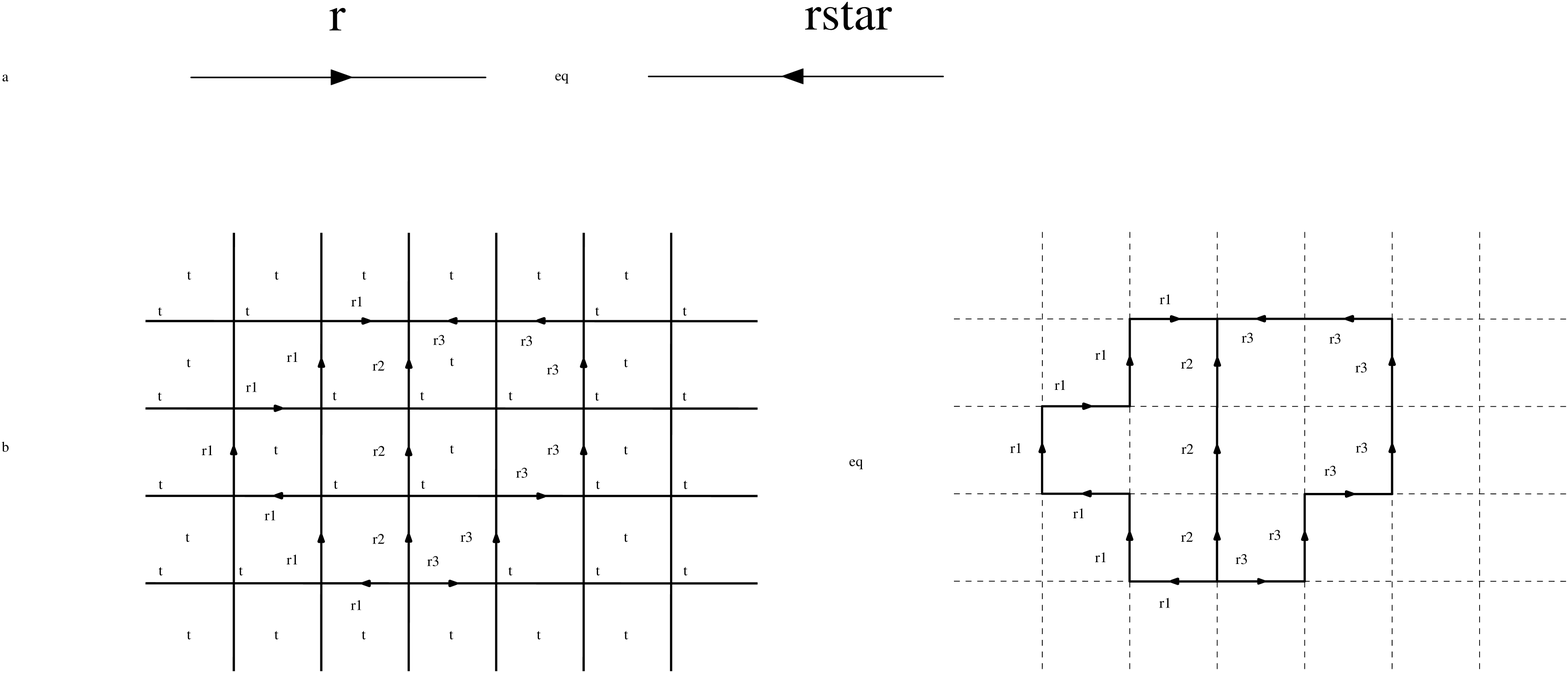}
\caption{\label{equivalenceofspinnetworks} Equivalence of spin networks.}
\end{figure}

\psfrag{eq}{$=$}
\psfrag{r}{$\rho$}
\psfrag{rstar}{$\rho^*$}
\psfrag{t}{$\rm\ssst triv$}
So far, the correspondence between spin networks and spin network states is not one-to-one, since many different spin networks yield the same functional. It follows from the definition of the dual representation $\rho^*$ on $V_{\rho^*} \equiv V^*_\rho$, for instance, that
\be
I_{\ssst 2}{}^{\cdots}{}_{\cdots a}\; \rho_e(g^{-1}_e)^a{}_b\; I_{\ssst 1}{}^{b\,\cdots}{}_{\cdots}
= I_{\ssst 1}{}^{b\,\cdots}{}_{\cdots}\; \rho^*_e(g_e)_b{}^a\; I_{\ssst 2}{}^{\cdots}{}_{\cdots a}\,.
\ee
This means that we can reverse a spin network edge $e$, change the label from $\rho_e$ to $\rho^*_e$, and still obtain the same spin network state (\fig{equivalenceofspinnetworks}a). Likewise, a spin network with trivially labelled edges defines the same functional as a spin network where these edges have been removed: representation tensors of the trivial irrep contribute just factors of 1 (\fig{equivalenceofspinnetworks}b).

We take account of this redundancy and consider spin networks as {\it equivalent} when they define identical spin network functionals.

The use of the spin network states lies in the fact that they span the space $\clL^2_0(\clA_\kappa)$ of gauge-invariant functionals of the lattice connection. To proof this, one has to apply the Peter-Weyl theorem to each edge of $\kappa$ \cite{Baezspinnetworkstatesingaugetheory}. Moreover, if we select orthonormal bases of intertwiners for the spaces \eq{tensorproduct}, and take only these basis tensors as labels for spin networks, the resulting states form an orthonormal basis of $\clL^2_0(\clA_\kappa)$. We call this basis $B_\kappa$.

By sum over $S_f$ in $B_f$ we mean that we sum over all spin networks $S_f$ whose {\it states} $\Psi_{S_f}$ are in the basis $B_f$. 

Analogously, we construct orthonormal bases for functionals over subgraphs of $\kappa$: for example, when we consider only spin networks on the boundary $\pa\kappa$, we obtain an orthonormal basis $B_{\pa\kappa}$ of $\clL^2_0(\clA_{\pa\kappa})$. The smallest admissible graphs consist of edges that surround a face $f$. The spin networks on such a graph are loops and provide an orthonormal basis $B_f$ for functionals like the face action \eq{Wilsonaction}, which depend only on the connection around $f$.

\psfrag{f}{$\sst f$}
Consider a loop spin network in $B_f$ whose edges are coherently oriented as in \fig{aloopspinnetwork}. Then, all intertwiners are of the form $I_v\in V_{\rho_1}\otimes V^*_{\rho_2}$. Due to Schur's lemma, this intertwiner is only non-zero if $\rho_1 = \rho_2 = \rho$, and in that case, it is a multiple of the identity. Thus, we have only a single basis intertwiner in $V_\rho\otimes V^*_\rho$, and since it is normalized, it is fixed up to the choice of a phase: we take the phase factor to be one, so that
\be
I^a{}_b = \frac{1}{(\dim V_\rho)^{1/2}}\,\delta^a{}_b\,.
\label{choiceoftrivialintertwiner}
\ee
When we insert this into the spin network functional, factors of dimension from edges and intertwiners cancel each other and we get \setlength{\jot}{0.2cm}
\bea
\Psi_{S_f}(g) &=& \tr\left[\rho(g^{-1}_{e_4})\rho(g^{-1}_{e_3})\rho(g_{e_2})\rho(g_{e_1})\right] \\
&=& \tr\left[\rho(U_f(g))\right]\,.
\label{loopfunctional}
\eea
That is, the loop functional is just the trace of the face holonomy in the irreducible representation $\rho$.
\pic{aloopspinnetwork}{A loop spin network.}{4cm}{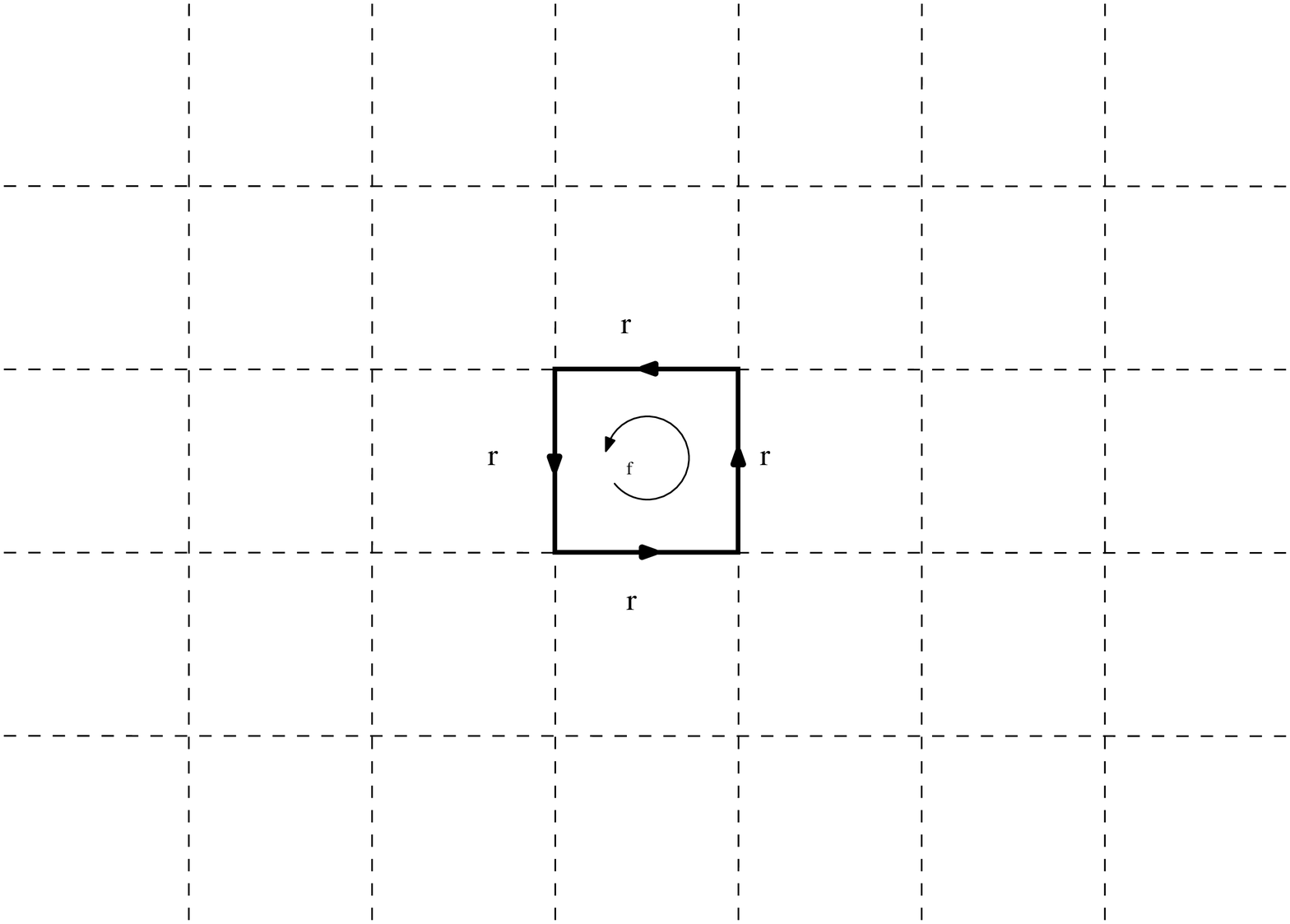}
\setlength{\jot}{0cm}

\section{Dual transformation to the spin foam model}
\label{dualtransformationtothespinfoammodel}

Let us now come to the actual dual transformation: for a given ultralocal lattice gauge theory, it maps the functional integral \eq{generalamplitude} over connections into a discrete sum over so-called {\it spin foams}: two-dimensional surfaces with branchings that have a certain labelling by irreps and intertwiners. Each spin foam is weighted by a factor---the spin foam amplitude---which is the analogue of the action and measure factors in the original theory. We can apply this transformation for every observable $\Phi$, and thus obtain a new, but physically equivalent formulation of the gauge theory: connections are replaced by spin foams, action and measure by the spin foam amplitude, and observables $\Phi$ are translated into spin foam dependent weighting factors $\tilde{\Phi}$. We call this new theory a {\it spin foam model}.

The dual transformation of the path integral is achieved by expanding all functionals into spin network states, and by a subsequent integration over the connection. Here, we restrict ourselves to the case, where the amplitude is of the form \eq{boundaryamplitude}. The procedure for more general amplitudes \eq{generalamplitude} is analogous.

It is convenient to split the evalutation of
\be
\Omega(\Phi) = \int\left({\ts\prod\limits_{e\in\kappa}}\d g_e\right)\;\exp\left(\iota \clS(g)\right)\Phi^*(\pa g)\,.
\label{originalpathintegral}
\ee
into two steps. At first, we integrate the exponentiated action over group elements on the interior $\kappa^\circ$: this leaves the connection on the boundary as a free variable, so we obtain a functional
\be
\Omega(g) := \int\limits_{\pa g' = g}\!\!\!\left({\ts\prod\limits_{e\in\kappa^\circ}}\d g'_e\right)\;\exp\left(\iota \clS(g')\right)\,.
\label{functionalOmega} 
\ee
In the second step, we convolute this functional with the boundary state $\Phi$, which yields the complete amplitude
\be
\Omega(\Phi) =  \int\left({\ts\prod\limits_{e\in\pa\kappa}}\d g_e\right)\;\Omega(g)\Phi^*(g)\,.
\label{contractionofOmegawithboundarystate}
\ee
Recall that the action decomposes into face actions, so \eq{functionalOmega} can be rewritten as
\be
\Omega(g) = \int\limits_{\pa g' = g}\!\!\!\left({\ts\prod\limits_{e\in\kappa^\circ}}\d g'_e\right)\,\prod_{f\in\kappa^\circ}\exp\left(\iota \clS_f(g')\right)\,.
\label{functionalOmegaafterusingultralocalproperty}
\ee
The expansion of $\exp\left(\iota \clS_f(g)\right)$ into the spin network basis $B_f$ gives
\be
\exp(\iota \clS_f) = \sum_{S_f \in B_f} c_{S_f}\Psi_{S_f}\,.
\label{expansionofplaquetteaction}
\ee
The subscript $S_f\in B_f$ means that we sum over all spin networks $S_f$ whose states  $\Psi_{S_f}$ lie in the basis $B_f$. We see from this and equation \eq{loopfunctional} that both action and spin network states depend only on the face holonomy $U_f$, and that \eq{expansionofplaquetteaction} is just another way of writing the character expansion
\be
\exp\left(\iota \clS_f(U_f)\right) = \sum_\rho \dim V_\rho\,c_{f\rho}\,\chi_\rho(U_f)\,.
\label{characterexpansion}
\ee
In the following, we assume that for the trivial representation the coefficients $c_{f\rho}$ are 1. Otherwise we redefine the coefficients suitably.

\psfrag{arrow}{$\rightarrow$}
\psfrag{brline}{\small branching line}
\psfrag{ubrcomp}{\small unbranched component}
By plugging \eq{expansionofplaquetteaction} into the functional \eq{functionalOmegaafterusingultralocalproperty}, we get
\bea
\Omega(g) &=&
\int\limits_{\pa g' = g}\!\!\!\left({\ts\prod\limits_{e\in\kappa^\circ}}\d g'_e\right)\,\prod_{f\in\kappa^\circ}\sum_{S_f \in B_f} c_{S_f}\Psi_{S_f}(g) \\
&=&
\sum_{\{f\}\to \{S_f\}}\;\,\int\limits_{\pa g' = g}\!\!\!\left({\ts\prod\limits_{e\in\kappa^\circ}}\d g'_e\right)
\prod_{f\in\kappa^\circ} c_{S_f}\Psi_{S_f}(g)\,.
\label{beforeintegratingout}
\eea 
In the second line, we pulled the summation symbol to the front, so we have to sum over all possible ways to assign basis spin networks $S_f$ to the faces $f$ of $\kappa^\circ$.

Consider a single term in this sum, i.e.\ a summand for a fixed choice of basis spin network $S_f$ for each face:
\be
\int\limits_{\pa g' = g}\!\!\!\left({\ts\prod\limits_{e\in\kappa^\circ}}\d g'_e\right)
\prod_{f\in\kappa^\circ} c_{S_f}\Psi_{S_f}(g')
\label{termforfixedchoice}
\ee
To determine the value of this term, we organize the appearing non-trivial loop spin networks into surfaces: two non-trivial loops belong to the same surface if they have only one common link and do not share this link with a third non-trivial loop. Different surfaces are either disconnected or they meet along links with other surfaces. The intersection of surfaces defines a graph which we call the {\it branching graph}: each line of this graph\footnote{We refer to lines of the branching graph also as links, but one should not confuse them with links or edges of the lattice. A single link of the branching graph can be built from many lattice links.} must join at least three surfaces, otherwise their loops would form a single surface. Loops outside of surfaces carry the trivial representation. The ensemble of surfaces can be viewed as a {\it branched surface} $\clF$ that consists of {\it unbranched components} $\clF_i$ (see \fig{groupingofloopsintounbranchedcomponents}). We denote the branching graph by $\Gamma_\clF$.
\pic{groupingofloopsintounbranchedcomponents}{Grouping of loops into unbranched components.}{5cm}{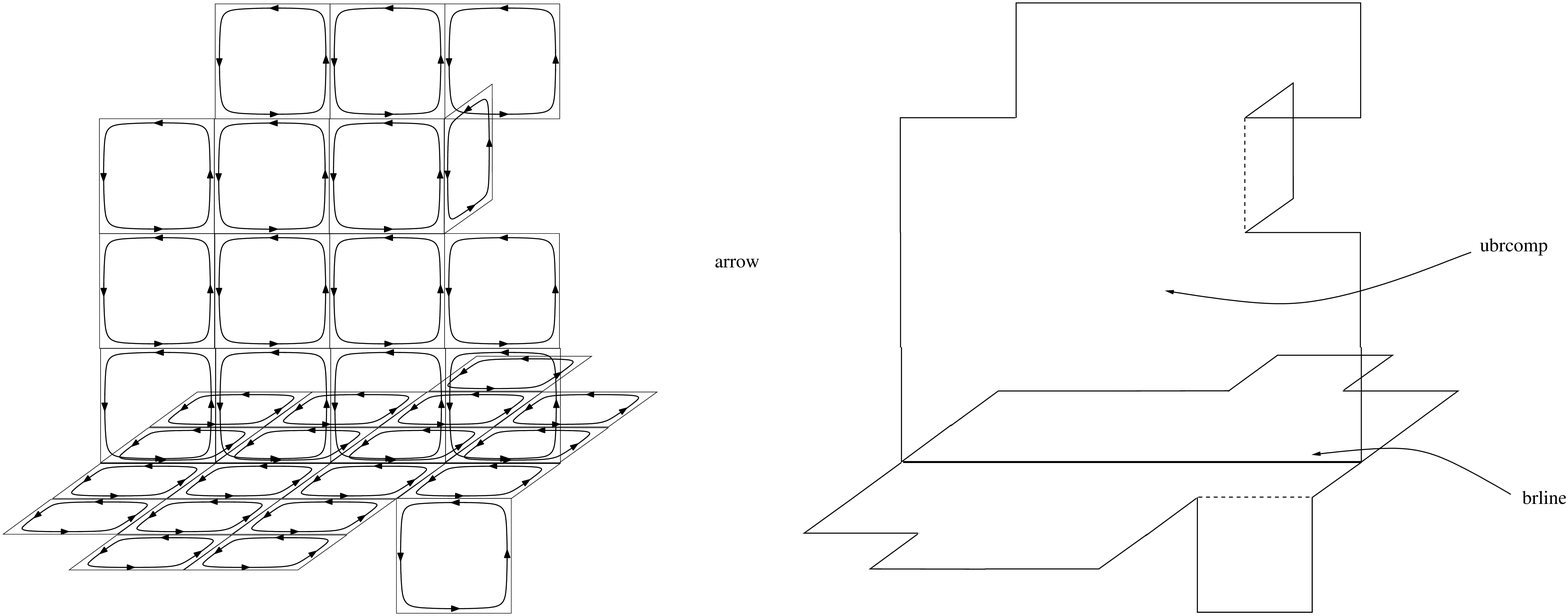}

\psfrag{r}{$\sst\rho$}
Without loss of generality, we can choose representatives of loops that have coherent orientations within each
unbranched component: i.e.\ orientations as in diagrams on Stoke's theorem.
To represent the group integration, we use the diagrammatics of \cite{GirelliOecklPerez}: loop edges with label $\rho_e$ symbolize the representation maps $\rho_e(g_e)$ and integrations over group elements are indicated by ``cables'' that surround the edges (see \fig{representationofgroupintegrationsbycables}).
\pic{representationofgroupintegrationsbycables}{Representation of group integrations by cables (irrep and intertwiner labels are omitted).}{5cm}{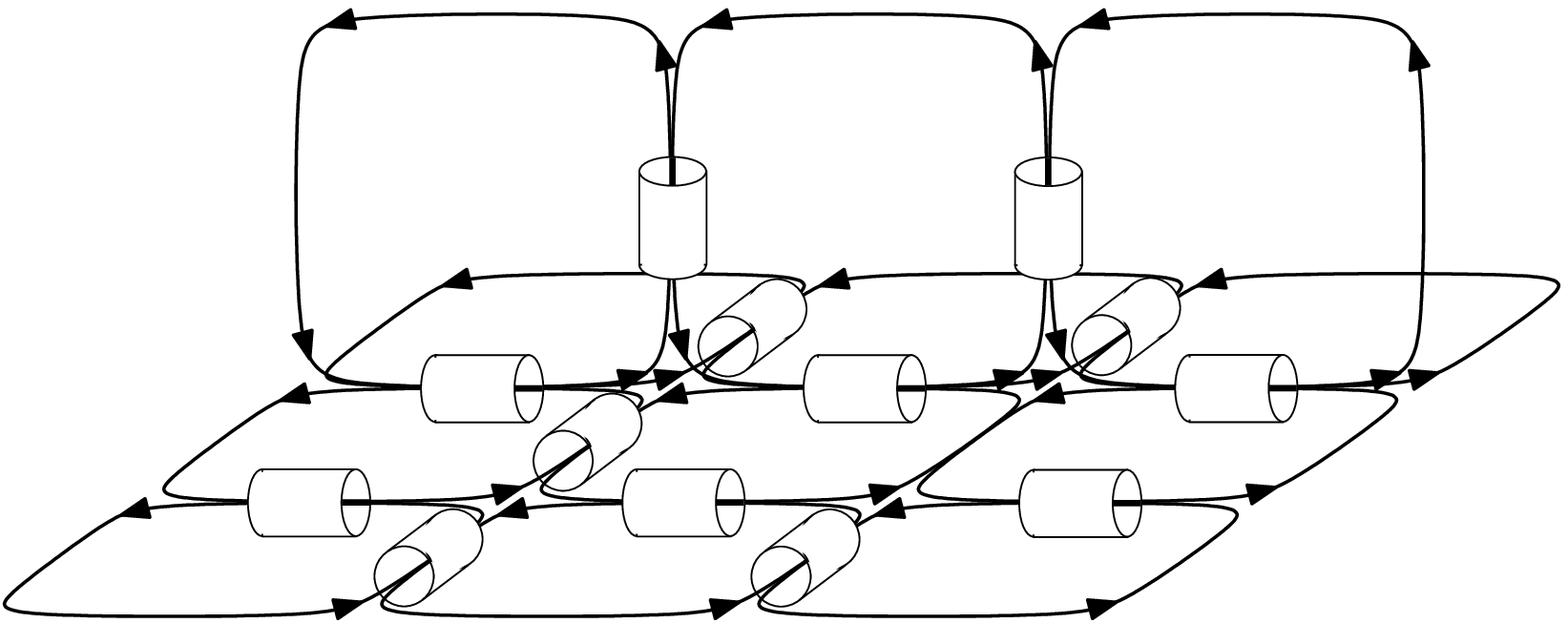}

\psfrag{e}{$e$}
The next step consists in integrating out the group variables on each lattice edge $e$. When doing so, we have to distinguish between four different cases:
\begin{enumerate}
\item All incident loops on $e$ are trivially labelled. In that case, the representation maps $\rho_e(g_e)$ are factors of 1 and the normalized Haar measure yields 1 as well. Therefore, all parts of the diagram that lie outside the branched surface contribute just 1 and disappear from the calculation.
\item Only one non-trivial loop is incident on $e$. Then, it follows from the identity
\be
\int_G \d g\;\, \rho_1(g)^{a_1}{}_{b_1}\,\rho_2(g^{-1})^{b_2}{}_{a_2}
= \frac{1}{\dim V_{\rho_1}}\,\delta{}^{a_1}{}_{a_2}\,\delta\,{}_{b_1}{}^{b_2}\,\delta_{\rho_1\,\rho_2}\,.
\label{orthogonalityrelation}
\ee
that the integral is zero, since $\rho\neq {\rm\sst triv}$. This implies that an unbranched component is not allowed to end in a trivial part of the diagram, otherwise it gives zero in \eq{termforfixedchoice}.
\item When exactly two non-trivial loops are incident on $e$, the representation labels have to be the same, due to \eq{orthogonalityrelation}. Thus, an unbranched component can only consist of loops with a single representation. We say that it is single-colored.
\item When $e$ is shared by more than two non-trivial loops, the integral over $G$ yields a so-called Haar intertwiner $H$. For a cable as in \fig{exampleofacable}, we obtain, for example,
\bea
H^{a_1a_2}{}_{a_3\,b_1b_2}{}^{b_3} &=& \int_G\d g\; \rho_1(g)^{a_1}{}_{b_1}\,\rho_2(g)^{a_2}{}_{b_3}\,\rho_3(g^{-1})^{b_3}{}_{a_3} \\
&=& \int_G\d g\; \rho_1(g)^{a_1}{}_{b_1}\,\rho_2(g)^{a_2}{}_{b_3}\,\rho^*_3(g)_{a_3}{}^{b_3}\,.
\eea
It is easy to see that tensors of this type are projectors onto invariant subspaces. In this example, $H$ is a projector onto the invariant part of $V_{\rho_1}\otimes V_{\rho_2}\otimes V^*_{\rho_3}$.
\pic{exampleofacable}{Example of a cable.}{2cm}{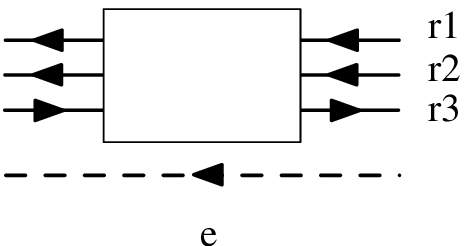}
\end{enumerate}
Our reasoning shows that in \eq{beforeintegratingout} we only need to sum over certain configurations of loops---configurations that can be regarded as branched colored surfaces: each unbranched component is built from faces in the interior of the lattice and carries a non-trivial representation label or color. It also comes with an orientation that is determined by the coherent orientation of loops we started from. When an unbranched component has a boundary, the boundary links have to lie on the lattice boundary or on the branching graph where more than two components meet. An ``open ending'' is not allowed due to point 3.

\psfrag{dr1r2}{$\sst\delta_{\rho_1\!,\rho_2}$}
\psfrag{r1}{$\sst\rho_1$}
\psfrag{r2}{$\sst\rho_2$}
\psfrag{rnm1}{$\sst\rho_{n-1}$}
\psfrag{rn}{$\sst\rho_n$}
\psfrag{dts}{$\vdots$}
\psfrag{sumi}{$\sum\limits_i$}
\psfrag{Ii}{$I_i$}
\psfrag{Iistar}{$I^*_i$}
\psfrag{delta}{$\delta$}
\psfrag{rho}{$\rho$}
To evaluate the contribution from a branched surface, we need to collect all factors and tensors that remain after the group integrations. We do this with the help of a graphical calculus that is close to that of \cite{GirelliOecklPerez}. Tensors and their contractions are represented by oriented labelled graphs: every vertex in the graph stands for a tensor $T_v$ (indicated by a label $T_v$), while the edges between vertices symbolize the contraction of tensor indices. As in spin networks, the edges carry irreducible representations as labels. The irrep specifies the representation space for which the contraction is performed. At a vertex, the irreps of incoming and outgoing edges determine in which tensor space the tensor lives (cf.\ \eq{tensorproduct} and \fig{spinnetwork}b). 
When an edge forms a closed curve, we interpret it as the trace in the respective representation.
Note that two edges at a delta tensor are equivalent to a single edge without vertex:
\be
\parbox{7cm}{\includegraphics[height=1cm]{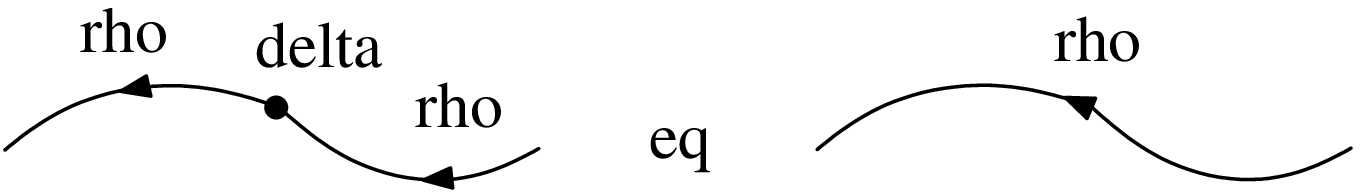}}
\ee 
With these conventions\footnote{Observe the difference to the rule for spin network functionals: in the case of spin networks, both vertices and edges are translated into tensors, while here the edges stand exlusively for contractions.}, the identity \eq{orthogonalityrelation} takes the form
\be
\parbox{10cm}{\includegraphics[height=0.9cm]{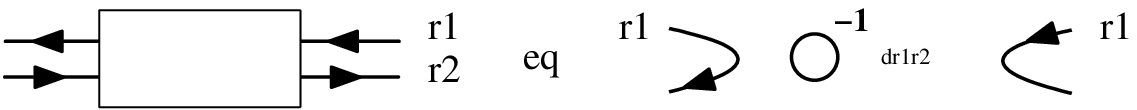}}\,,
\label{cableidentitywithtwostrands}
\ee
By ``circle to the minus one'' we mean the inverse of the dimension.

The ``splitting identity'' \eq{cableidentitywithtwostrands} is just a special case of that for the Haar intertwiner
\be
H^{a_1\cdots a_n}{}_{b_1\cdots b_n} = \int_G \d g\;\rho_1(g)^{a_1}{}_{b_1}\cdots \rho_n(g)^{a_n}{}_{b_n}\,.
\ee
in which case
\be
\parbox{9.5cm}{\includegraphics[height=2cm]{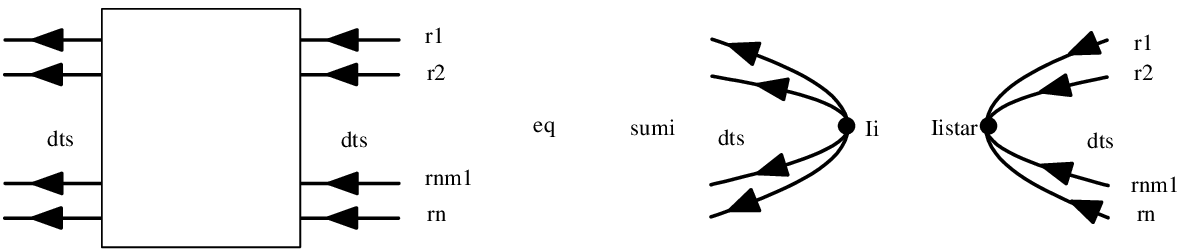}}\,.
\label{cableidentitywithnstrands}
\ee
The two diagrams are equal because $H$ is a projector and can be written as a sum over basis intertwiners $I_i$:
\be
H^{a_1\cdots a_n}{}_{b_1\cdots b_n} = \sum_i\, I_i{}^{a_1\cdots a_n}\,I_i\,{}_{b_1\cdots b_n}\,.
\label{Haarintertwineridentity}
\ee

Let us start by considering a single-colored component $\clF_i$. We see from diagram \ref{integrationsonasinglecoloredcomponent} that we get the dimension $\dim V_\rho$ for each vertex in the interior of $\clF_i$, and a factor $(\dim V_\rho)^{-1}$ for each interior edge. Each face contributes the loop coefficient
\be
c_{S_f} = \dim V_\rho\,c_{f\rho}
\ee
(cf.\ eqns.\ \eq{expansionofplaquetteaction} and \eq{characterexpansion}). When the unbranched component is without boundary, the factors accumulate to the amplitude
\be
A_{\clF_i} = (\dim V_\rho)^{\chi(\clF_i)}\,\prod_{f\in \clF_i} c_{f\rho}\,.
\label{amplitudeofunbranchedcomponent}
\ee
Here, $\chi(\clF_i)$ stands for the Euler number of the unbranched component.
\pic{integrationsonasinglecoloredcomponent}{Integrations on a single-colored component.}{6cm}{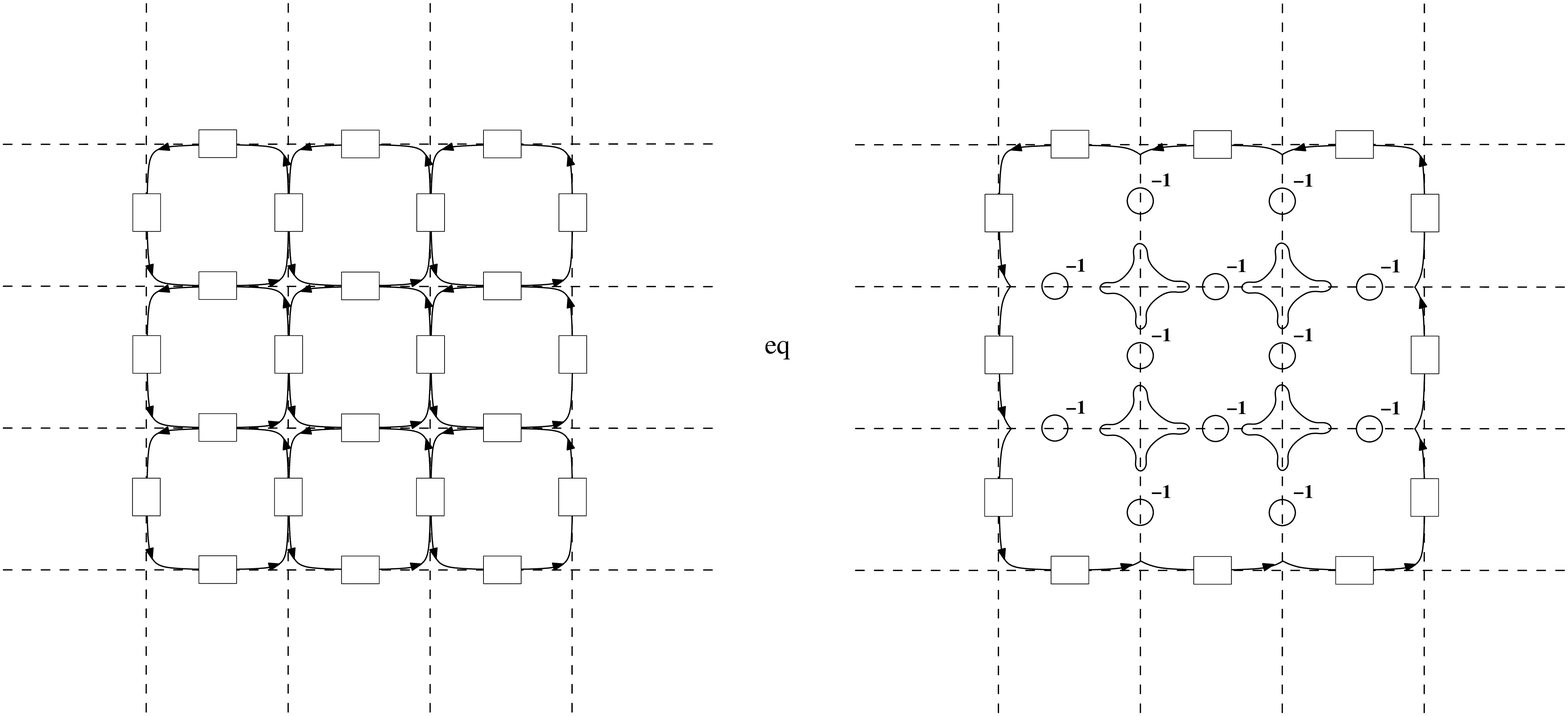}

This formula is still true when $\clF_i$ has a boundary. Since $\clF_i$ is 2-dimensional, its boundary can only consist of components that have the topology of a circle. As shown in \fig{integrationsonasinglecoloredcomponent}, the vertices and edges on the boundary fail to produce any dimensional factors. The missing factors, however, would just add up to
\be
(\dim V_\rho)^{\chi(S^1)} = 1\,.
\ee
Therefore, equation \eq{amplitudeofunbranchedcomponent} remains valid.

\psfrag{H}{$H$}
Next we analyze the contribution from the branching graph $\Gamma_\clF$: as explained in point 4., each lattice edge on a branching line produces a Haar intertwiner $H$. Between two vertices of $\Gamma_\clF$, all Haar intertwiners are the same, because the irrep labelling of incident loops does not change. Thus, we can use the projection property of $H$ to replace the sequence of $H$ factors by a single one (see \fig{Haarintertwinersonbranchingline}).
\pic{Haarintertwinersonbranchingline}{Haar intertwiners on branching line.}{6cm}{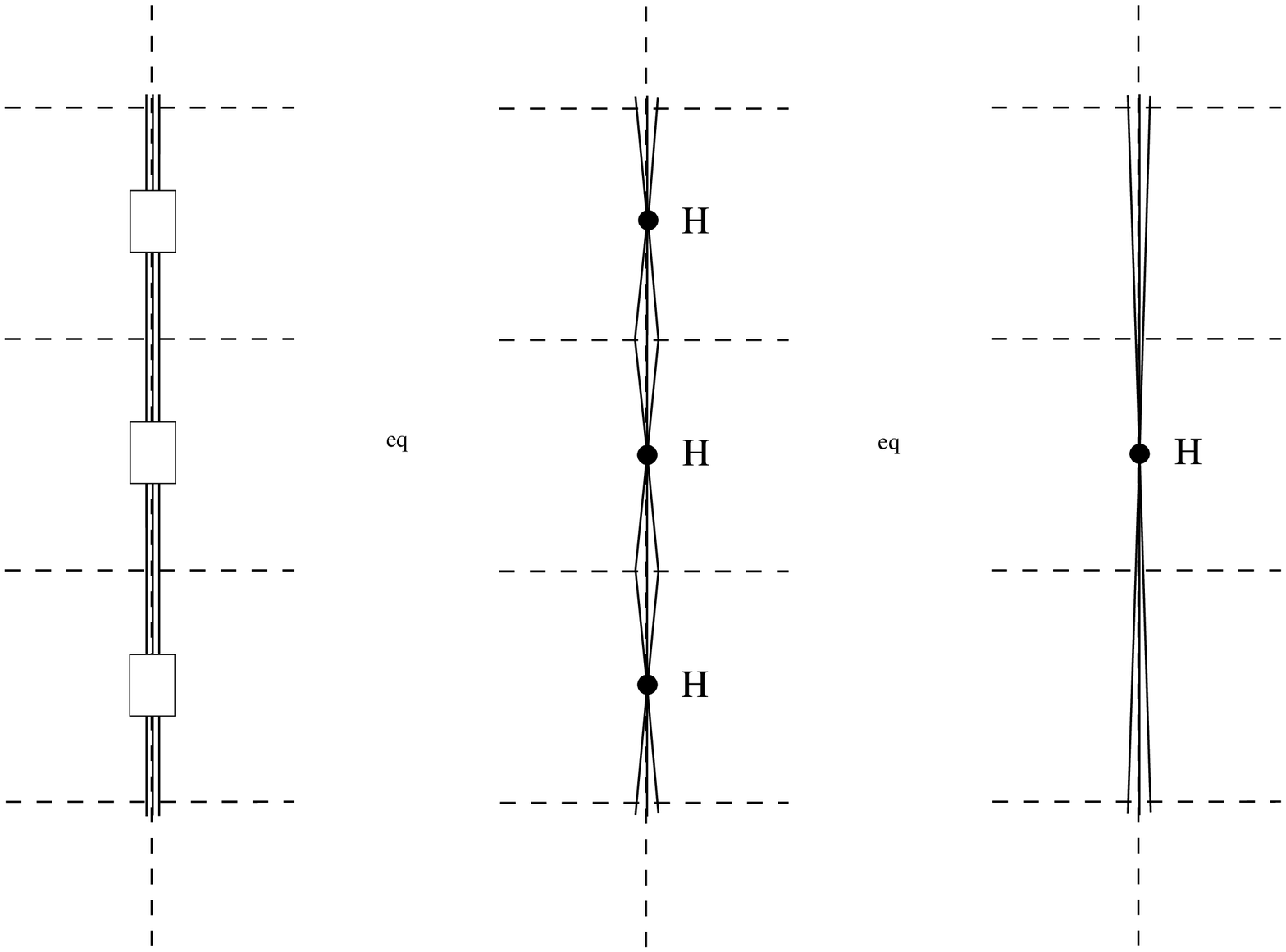}

\psfrag{H1}{$H_1$}
\psfrag{H2}{$H_2$}
\psfrag{H3}{$H_3$}
\psfrag{H4}{$H_4$}
\psfrag{Ii1}{$I_{i_1}$}
\psfrag{Ii2}{$I_{i_2}$}
\psfrag{Ii3}{$I_{i_3}$}
\psfrag{Ii4}{$I_{i_4}$}
\psfrag{Ii1star}{$I^*_{i_1}$}
\psfrag{Ii2star}{$I^*_{i_2}$}
\psfrag{Ii3star}{$I^*_{i_3}$}
\psfrag{Ii4star}{$I^*_{i_4}$}
\psfrag{sumi1}{$\sum_{i_1}$}
\psfrag{sumi2}{$\sum_{i_2}$}
\psfrag{sumi3}{$\sum_{i_3}$}
\psfrag{sumi4}{$\sum_{i_4}$}
\psfrag{Ie1}{$I_{l_1}$}
\psfrag{Ie2}{$I_{l_2}$}
\psfrag{Ie3}{$I_{l_3}$}
\psfrag{Ie4}{$I_{l_4}$}
\psfrag{Ie5}{$I_{l_5}$}
\psfrag{dots}{$\ldots$}
\psfrag{r12}{$\sst\rho_{12}$}
\psfrag{r13}{$\sst\rho_{13}$}
\psfrag{r14}{$\sst\rho_{14}$}
\psfrag{r15}{$\sst\rho_{15}$}
\psfrag{r23}{$\sst\rho_{23}$}
\psfrag{r24}{$\sst\rho_{24}$}
\psfrag{r25}{$\sst\rho_{25}$}
\psfrag{r34}{$\sst\rho_{34}$}
\psfrag{r35}{$\sst\rho_{35}$}
\psfrag{r45}{$\sst\rho_{45}$}
After having done this for every branching line, we apply identity \eq{cableidentitywithnstrands}: each $H$ is split into a sum over products of two basis intertwiners, which we ``pull off'' towards the vertices of the branching graph (\fig{splittingofHaarintertwiners}). The result is a sum over possible assignments of basis intertwiners to links of $\Gamma_\clF$, and an amplitude
\be
A_v = \quad\parbox{5cm}{\includegraphics[height=4cm]{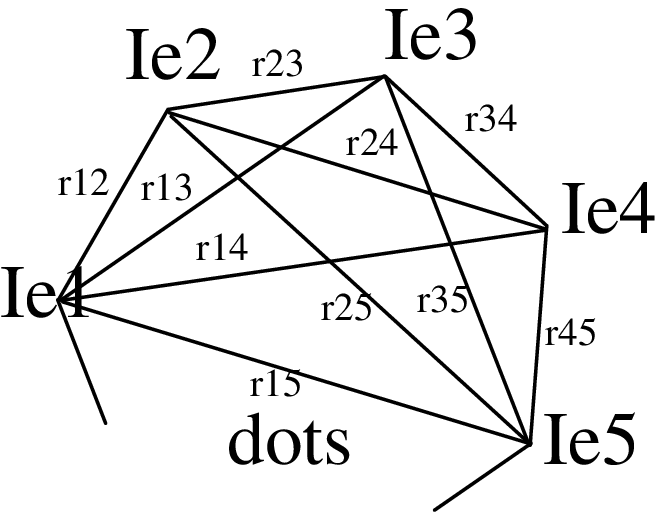}}
\label{amplitudeforvertexofbranchinggraph}
\ee
for each vertex of the branching graph. The intertwiners come from links $l_i$ of $\Gamma_\clF$ that are incident on $v$, and the irreps $\rho_{ij}$ belong to single-colored components that are bounded by pairs of links $l_i, l_j$.
\pic{splittingofHaarintertwiners}{Splitting of Haar intertwiners.}{4cm}{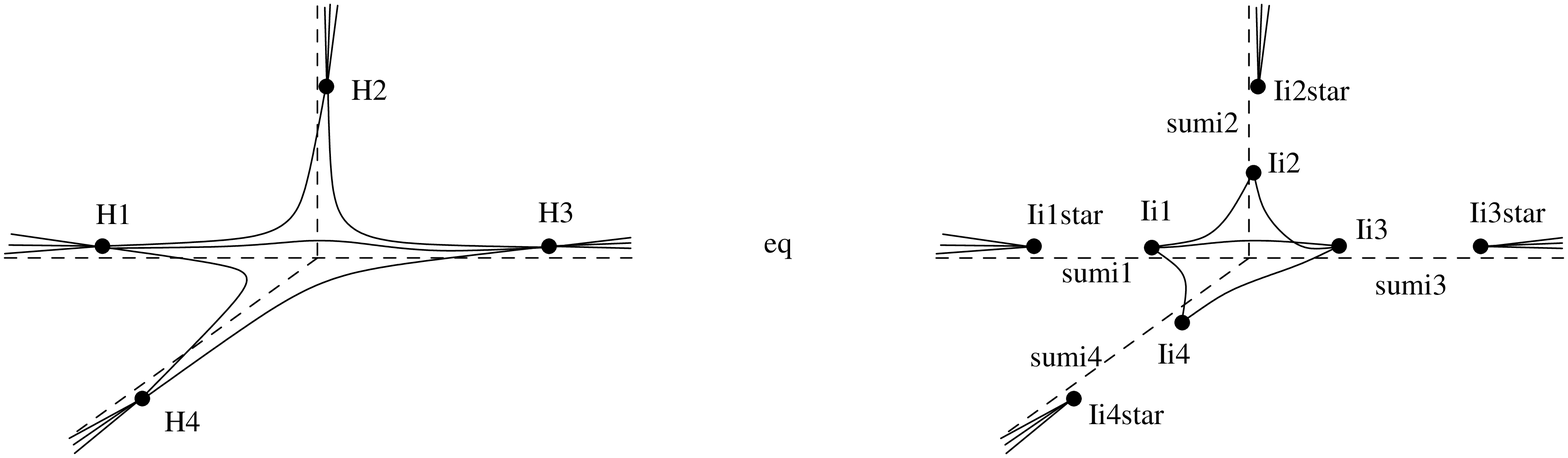}

\begin{figure}
\hspace{-1cm}\includegraphics[height=2.5cm]{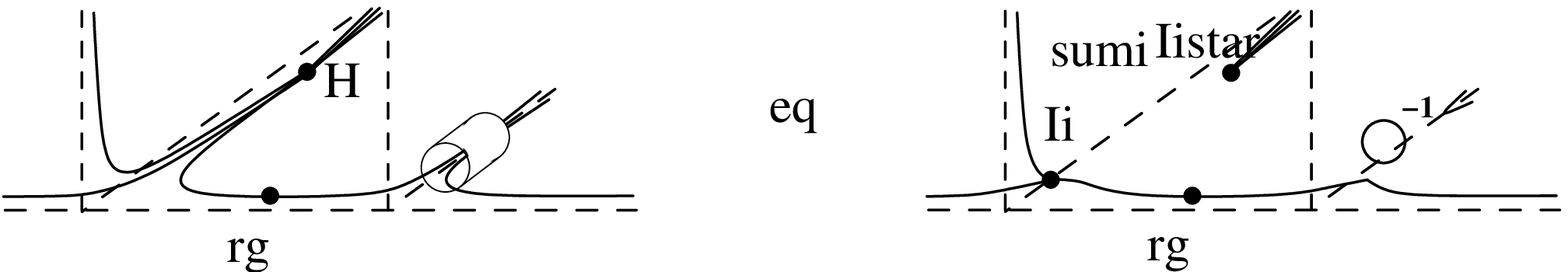}
\caption{\label{atlatticeboundary} Branching line and single-colored component at the lattice boundary. The unintegrated representation tensor $\rho_e(g_e)$ is symbolized by a vertex.}
\end{figure}

\psfrag{rg}{$\rho_e(g_e)$}
\psfrag{sumi}{$\sum_i$}
Additional contributions arise when branching lines and unbranched components connect to the lattice boundary $\pa\kappa$: as depicted in \fig{atlatticeboundary}, the splitting of Haar intertwiners ``pushes'' basis intertwiners out to the lattice boundary and the representation tensors $\rho_e(g_e)$ of boundary edges are left unintegrated. Therefore, what we get on the boundary is exactly a spin network functional! The associated spin network is determined by the branched surface $\clF$ and can be visualized as a cut of the boundary through $\clF$ (see \fig{spinfoamsasworldsheetofspinnetworks}): its intertwiners on vertices are fixed by the intertwiners on branching lines of $\clF$, and the colors on edges equal the colors on unbranched components. We denote the boundary spin network by $S_F$ and its functional by $\Psi_{S_\clF}$.

For later convenience, we want $\Psi_{S_\clF}$ to be normalized: this is not yet the case, because the 2-valent intertwiners are just $\delta$'s and not $\delta / (\dim V_\rho)^{1/2}$, and the edges do not carry any factors $(\dim V_\rho)^{1/2}$. We add the needed factors to $\Psi_{S_\clF}$, and compensate this by absorbing their inverse into the amplitude of the single-colored surfaces $\clF_i$. The precise form of the correction factor depends on the topology of $\clF_i$ : let $B_{ij}$ denote the components of the boundary $\pa F_i$ that lie on the lattice boundary $\pa\kappa$.
For each component $B_{ij}$ that is an open line, we receive a correction factor $(\dim V_\rho)^{-1/2}$, while we get just 1 when it is a loop. Thus, we can express the corrected amplitude for $\clF_i$ as
\be
A_{\clF_i} = (\dim V_\rho)^{\tilde{\chi}(\clF_i)}\,\prod_{f\in \clF_i} c_{f\rho}
\label{correctedamplitudeforFi}
\ee
where
\be
\tilde{\chi}(\clF_i) = \chi(\clF_i) - \frac{1}{2}\sum_j \chi(B_{ij})\,.
\ee

\psfrag{I}{$I$}
\psfrag{Istar}{$I^*$}
\psfrag{GF}{$\Gamma_\clF$}
\psfrag{paM}{$\pa M$}
For sake of completeness, we should mention that there are degenerate examples of branching graphs which we disregarded so far: a line of the branching graph can be closed or go directly from boundary to boundary (see \fig{degenerateexamplesofbranchinggraphs}). In that case, there is no vertex on the branching line, but nevertheless the splitting of Haar intertwiners leads to a diagram. It has the particularly simple form
\be
\parbox{4cm}{\includegraphics[height=3cm]{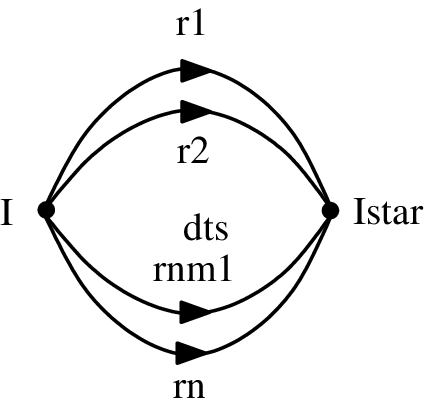}}\,.
\ee
and just gives 1, since the basis intertwiners are normalized.
\pic{degenerateexamplesofbranchinggraphs}{Degenerate examples of branching graphs.}{2cm}{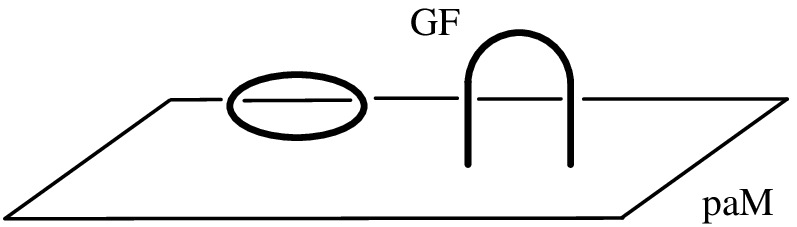}

Let us summarize what we found: the functional $\Omega(g)$ can be written as a sum over terms which are characterized by branched surfaces $\clF$ with a labelling: each unbranched component $\clF_i$ of $\clF$ carries an orientation and an irrep label, and every branching line comes with an intertwiner.

We call such a surface $\clF$, together with its labelling, a {\it spin foam} on the lattice $\kappa$.
Remember that the initial choice of orientation on unbranched components was arbitrary. With the opposite sense, the representation label would be the dual, but everything else, including amplitudes, would remain unchanged. For that reason, we identify spin foams that are related by a reorientation and dualization of a single-colored component.

Given this definition, the functional $\Omega(g)$ is equal to a sum over all spin foams $\clF$ on $\kappa$, where each spin foam contributes a spin network functional $\Psi_{S_\clF}$ and an amplitude factor:
\be
\Omega(g) = \sum_{\clF\subset\kappa}\left(\prod_{v\in\Gamma_\clF} A_v\right)\left(\prod_i A_{\clF_i}\right)\,\Psi_{S_\clF}(g)\,.
\label{Omegaasaspinfoamsum}
\ee
For each vertex of $\Gamma_F$, we get a factor of the form \eq{amplitudeforvertexofbranchinggraph} and each single-colored component yields the amplitude \eq{correctedamplitudeforFi}.

Note that the amplitude of the branching graph is geometric in the sense that it only depends on the connectivity and labelling of the graph, but not on how its lines are discretized by edges of the lattice. The amplitude for single-colored components $\clF_i$ contains also a geometric part, given by
\be
(\dim V_\rho)^{\tilde{\chi}(\clF_i)}
\ee
which makes no reference to the lattice. In general, the product
\be
\prod_{f\in \clF_i} c_{f\rho}
\label{productoffacefactors}
\ee
of face factors {\it does} depend on the discretization and constitutes the non-geometric part of the spin foam amplitude. It is only for special choices of $c_{f\rho}$ that the product \eq{productoffacefactors} has a geometric interpretation. An example for this is BF-theory (see below) and the model we present in the next section.

In the final step, we contract the functional $\Omega(g)$ with the boundary state $\Phi(g)$
to give the value of the complete path integral \eq{originalpathintegral}.
For that purpose, we expand $\Phi(g)$ in orthonormal basis spin networks:
\be
\Phi(g) = \sum_{S \in B(\pa\kappa)} \Phi_S\,\Psi_S(g)\,.
\label{expansionofboundarystate}
\ee
When contracting \eq{expansionofboundarystate} with \eq{Omegaasaspinfoamsum}, only those terms survive for which $S = S_\clF$, since $\Psi_{S_\clF}$ is a basis spin network. Therefore, the final result is
\be
\Omega(\Phi) = \sum_{\clF\subset\kappa}\left(\prod_{v\in\Gamma_\clF} A_v\right)\left(\prod_i A_{\clF_i}\right)\,\Phi^*_{S_\clF}\,.
\label{OmegaPsiasaspinfoamsum}
\ee
For each spin foam $\clF$, the amplitude is multiplied by the coefficent of the boundary state with respect to the spin network induced by $\clF$ on the boundary. Thus, we have attained a form for $\Omega(\Phi)$ that stands in analogy to the original path integral: the integration over connections has become a sum over spin foams, action and measure factors are replaced by the spin foam amplitude, and the weighting by the boundary functional $\Phi(\pa g)$ turns into the weighting by $\Phi_{S_\clF}$.

By the same procedure, we can also transform amplitudes \eq{generalamplitude} for other observables $\Phi$, and thereby translate all quantities of the lattice gauge theory into an equivalent theory that is purely formulated in terms of spin foams and spin networks. We refer to it as the spin foam model {\it dual} to the gauge theory.

The most simple spin foam model is that of BF theory (see eqns.\ \eq{plaquetteactionBFtheory} and \eq{formalpathintegralofBFtheory}). The plaquette action has the character expansion
\be
\delta(U_f) = \sum_\rho \dim V_\rho\,\chi_\rho(U_f)
\ee
so the plaquette coefficents $c_{f\rho}$ are all 1. The spin foam sum takes the form
\be
\Omega(\Phi) = \sum_{\clF\subset\kappa}\left(\prod_{v\in\Gamma_\clF} A_v\right)\left(\prod_i\left(\dim V_{\rho_i}\right)^{\tilde{\chi}(\clF_i)}\right)\Phi^*_{S_\clF}
\label{OmegaPsiBFtheory}
\ee

\subsection*{Geometric spin foams}

At this point, we should remark that the definition of spin foams that we employ here is not the same as the standard one appearing in the literature: usually, a spin foam is viewed as being built from {\it all} faces of the lattice, regardless of which labels they carry, and an amplitude is associated to every single vertex, edge and face. Here, we were led to merge sets of faces into more geometric objects that can be viewed as lying on the lattice (instead of {\it being} it), and the amplitudes of these objects depend only in part on the way they are subdivided by the lattice grid.

This organization of faces in ``larger surfaces'' was already emphasized in one of the first works on spin foams by Reisenberger \cite{ReisenbergerlatticefourdEuclidean}, and it was known before in lattice gauge theory (see e.g.\ \cite{DrouffeZuber}). The same idea is also contained in Baez' definition of a category of spin foams \cite{Baezspinfoammodels}.

To distinguish between the two notions of spin foam we call the ones defined here {\it geometric spin foams}:
we can think of them as branched labelled surfaces in the continuous manifold $M$ that are independent of any choice of lattice, similarly as a continuous field does not rely on a cutoff. From that point of view, the lattice regularization becomes the requirement that one should not sum over all possible geometric spin foams, but only over those that can be ``fitted'' onto the lattice $\kappa$: by that we mean that every single-colored component is a union of plaquettes of $\kappa$, and every branching line is a union of links of $\kappa$.

So far the lattice is just a means to regularize the theory, but it has no deeper conceptual justification. In the spin foam approach to gravity, one hopes to explain the spacetime lattice itself as being a consequence of spin foams. If that was true, we would arrive at a picture where spin foams of the gauge theory are placed on spin foams of gravity, and no external background structure or cutoff has to be imposed anymore. We will say more about this in section \ref{backgroundindependentspinfoammodels}.

\psfrag{SF}{$S_\clF$}
\psfrag{F}{$F$}
From a space {\it plus} time point of view, spin foams are interpreted as evolving spin networks \cite{ReisenbergerRovellisumoversurfaces}: when the boundary spin network consists of two disconnected spin network graphs, the interpolating spin foam can be viewed as the worldsheet of the transition from one spin network to the other (see \fig{spinfoamsasworldsheetofspinnetworks}). In the process, vertices can blow up to subgraphs, and subgraphs may shrink to a vertex. When there is only a single connected spin network, we regard the spin foam as a creation or annihiliation process.
\pic{spinfoamsasworldsheetofspinnetworks}{Spin foams as worldsheets of spin networks: the branching graph $\Gamma_F$ (thick line) indicates where surfaces with different irrep labels meet.}{4cm}{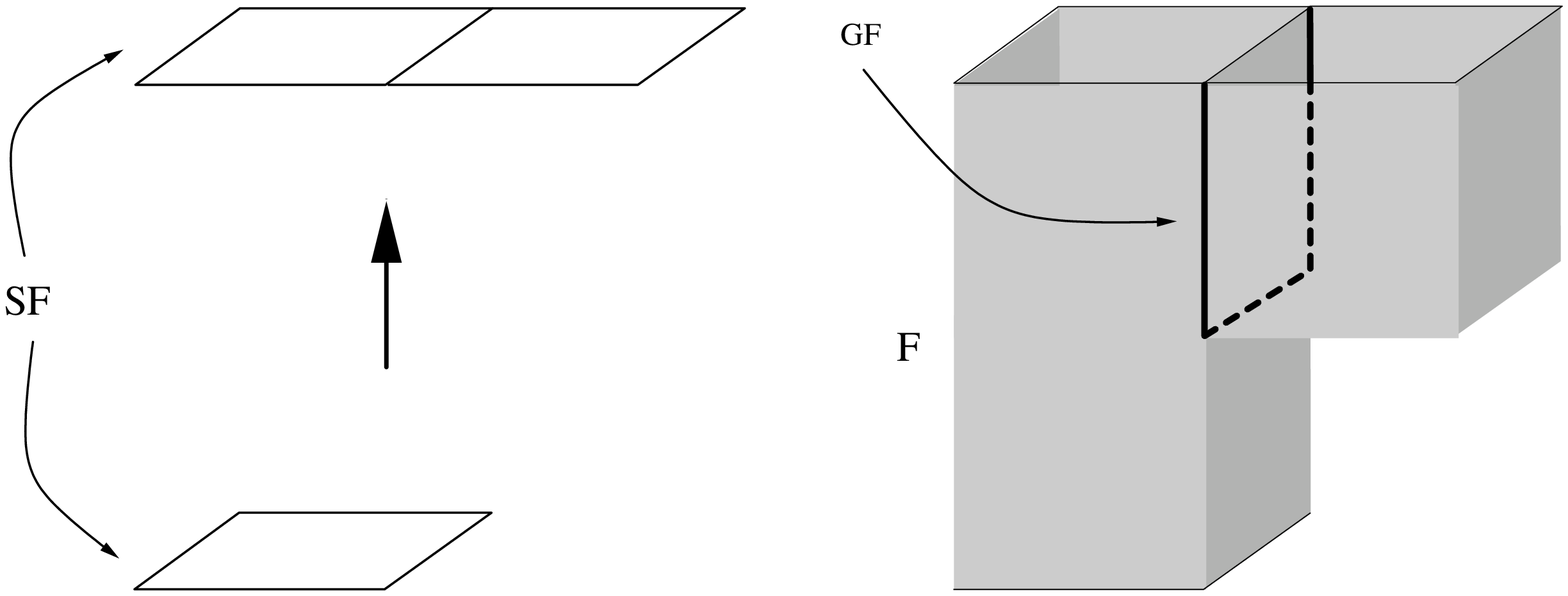}

\section{A spin foam model of Yang-Mills theory}
\label{aspinfoammodelofYangMillstheory}

In this section, we describe a spin foam model that is dual to Euclidean lattice Yang-Mills theory.
We take SU(N) as the gauge group. The spacetime dimension $d$ can have any value $\ge 2$.

As regards the choice of the face action, we have several possibilities, all of which produce the same classical continuum limit. The Wilson action \eq{Wilsonaction} appears as the most simple choice on the level of group variables, but under the dual transformation it leads to relatively complicated face coefficients $c_{f\rho}$.
In that respect, the most simple alternative is the heat kernel action, which has the character coefficients
\be
c_{f\rho} = \exp\left(- a^{(d-4)}\gamma^2(a)\,C_\rho\right)\,.
\label{cprhoheatkernalaction}
\ee
Here, $C_\rho$ is the eigenvalue of the Casimir operator in the $\rho$-representation.
On the side of the lattice field theory, this action takes the form
\be
\exp\left(- S_f(U_f)\right) = \frac{K\left(U_f,\frac{\gamma^2}{2}\right)}{K\left(\mathbbm{1},\frac{\gamma^2}{2}\right)}\,.
\label{heatkernelaction}
\ee
The heat kernel $K$ is determined by the differential equation
\be
\pdiff{}{t}K(U,t) = \Delta K(U,t)\,,\qquad K(U,0) = \delta(U)\,.
\ee
$\Delta$ denotes the Laplace-Beltrami operator on the group.

In the following, we take \eq{heatkernelaction} as the starting point for the dual transformation. The Wilson and heat kernel action are believed to be in the same universality class \cite{MontvayMuenster}, so the preference of one or the other should not affect the low-energy physics of the theory.

For a single-colored component $\clF_i$ with label $\rho_i$, the factors \eq{cprhoheatkernalaction} add up to
\be
\prod_{f\in \clF_i} c_{p\rho_i} = \exp\left(- N_{\clF_i}\,a^{(d-4)}\gamma^2(a)\,C_{\rho_i}\right)\,,
\ee
$N_{\clF_i}$ stands for the number of plaquettes in the surface $\clF_i$. Clearly, $N_{\clF_i}$ is proportional to the area $\clA_{\clF_i}$ of the single-colored component, so we can write
\be
N_{\clF_i} = \frac{\clA_{\clF_i}}{a^2}
\ee
and
\be
\prod_{f\in \clF_i} c_{p\rho_i} = \exp\left(- \clA_{\clF_i}\,a^{(d-6)}\gamma^2(a)\,C_{\rho_i}\right)\,.
\ee
The area is accompanied by factors which we absorb into the new constant
\be
T_\rho(a) := a^{(d-6)}\gamma^2(a)\,C_{\rho_i}\,.
\label{spinnetworktension}
\ee
The overall amplitude for a boundary state $\Phi$ becomes
\be
\Omega(\Phi) = \sum_{\clF\subset\kappa}\left(\prod_{v\in\Gamma_\clF} A_v\right)\left(\prod_i\left(\dim V_{\rho_i}\right)^{\tilde{\chi}(\clF_i)}\,\exp\Big(- T_{\rho_i} \clA_{\clF_i}\Big)\right)\Phi^*_{S_\clF}\,.
\label{OmegaPsiYangMills}
\ee
where the vertex amplitudes are defined as in equation \eq{amplitudeforvertexofbranchinggraph}.

As a result of the particular choice for the face coefficients (eqn.\ \eq{cprhoheatkernalaction}), we arrive at a model where the amplitude of a spin foam depends only on the geometric properties of the spin foam---its topology, labelling and area. The lattice regularization enters only through the value of $T_\rho(a)$ and the requirement that spin foams should be congruent with the lattice.

The form of the sum \eq{OmegaPsiYangMills} bears a striking resemblance to lattice quantizations of the Nambu-Goto string \cite{Dearnaley}: the spin foams play the role of worldsheets for the spin networks, and the exponent can be viewed as an action for each single-colored sheet that is proportional to its area. Thus, we interpret the factor \eq{spinnetworktension} as a color-dependent tension $T_\rho$ of spin network edges. The running of the original gauge coupling $\gamma(a)$ is mapped into a running of the edge tension.

\section{Background independent spin foam models}
\label{backgroundindependentspinfoammodels}

We have seen in the previous section that the notion of geometric spin foams allows for a remarkably simple dual formulation of Yang-Mills theory. In this section, we will argue that it is also particularly suited for constructing background-independent theories.

In the spin foam approach to gravity, one defines models that are no longer dual transforms of pure gauge theories.  In many cases, however, their definition is closely related to the duality map, as they can be obtained by modifiying the spin foam model dual to BF theory. The most prominent example is the Barrett-Crane model for Riemannian or Lorentzian gravity \cite{BarrettCrane}: for its heuristic derivation, we can start from a formal path integral over the tetrad and connection field. This path integral is rewritten as a BF path integral
\be
\int DA \int DB\; \exp\Big(\irm\,\tr\left[B F(A)\right]\Big)
\ee
with additional constraints on the $B$-field. One introduces a triangulation of spacetime, transforms pure BF theory into its dual spin foam model, and translates the constraints on $B$ into restrictions on the spin foams \cite{PerezspinfoamquantizationSOfourPlebanski}.

\psfrag{Ibc}{$I^{\ssst\rm BC}$}
In the case of the Riemannian Barrett-Crane model, the spin foam labels are restricted to balanced\footnote{A balanced representation is isomorphic to a tensor product $j\otimes j^*$ where $j$ is an irreducible represenation of SU(2).} representations of SO(4), and the intertwiners are replaced by so-called Barrett-Crane intertwiners $I^{\ssst\rm BC}$. If we denote such spin foams by $\FBC$, we can write the entire spin foam sum as
\be
\Omega(\Phi) = \sum_{\FBC\subset\kappa}\left(\prod_{v\in\Gamma_{\FBC}}
\parbox{2.65cm}{\includegraphics[height=2cm]{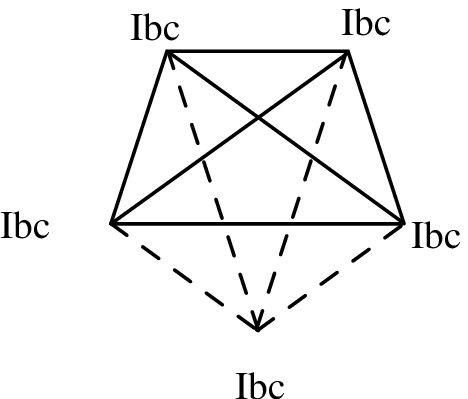}}
\right)\left(\prod_i\left(\dim V_{\rho_i}\right)^{\tilde{\chi}(\FBC{}_i)}\right)\Phi^*_{S_{\FBC}}\,.
\label{OmegaPsiBarrettCrane}
\ee
The lattice $\kappa$ consists of the dual complex of the chosen triangulation. The vertices of the branching graph have valence 4 or 5. Accordingly, the number of intertwiners in a vertex amplitude varies between 4 and 5 (indicated by dashed lines in \eq{OmegaPsiBarrettCrane}). 
We see that the transition from BF-theory to \eq{OmegaPsiBarrettCrane} preserved the geometric form of the sum---the spin foam amplitude has changed, but it is still a function of the geometry of the spin foam. 
There are other versions of the Barrett-Crane model that differ from \eq{OmegaPsiBarrettCrane} in that they contain additional modifications of the amplitude. In general, this breaks the topological invariance of surface amplitudes and makes our geometric interpretation impossible. We show in the appendix that \eq{OmegaPsiBarrettCrane} corresponds to version B in \cite{DePietriFreidelKrasnovRovelli}.

In either case---whether we regard spin foams as consisting of the entire lattice, or as geometric objects that are placed on it---the construction rests on the choice of a particular lattice $\kappa$, and that clashes with the idea of defining a theory that is independent of any background structure.

One way to resolve this problem goes along with the viewpoint that identifies spin foams with the lattice: the lattice itself is interpreted as a discrete spacetime, and the sum over spacetime fluctuations is then       implemented as a sum over a large class of lattices (and their labellings). Thus, by summing over lattices, one avoids the choice of any particular $\kappa$. The precise form of this sum is obtained from the perturbative expansion of a group field theory\footnote{a generalization of matrix theories} \cite{ReisenbergerRovelliconnectionformulation, ReisenbergerRovellispinfoamsasFeynmandiagrams}.

Here, we take a different approach, based on the philosophy that spin foams are lattice-independent geometric entities: we view the lattice just as an auxiliary background which was used for deriving the amplitudes, and discard it when defining the full background independent model. For this to be possible, we need to start from models whose amplitudes depend only on the geometry of spin foams, and not on the lattice on which the spin foam is defined. In the absence of a background metric, such amplitudes can only depend on topological properties---the topology of single-colored sheets and how they are connected by the branching line. 

Another way of stating this condition is to say that spin foam amplitudes must be topological invariants. Keep in mind that when we use the word ``topological'' in this way, it only refers to amplitudes of single spin foams, but it has nothing to do with topological invariance of the entire spin foam sum. The amplitudes of spin foams could be topological in the sense we just mentioned, and yet give a model that is not topologically invariant.

Consider a spin foam sum
\be
\Omega_\kappa(\Phi) = \sum_{\clF\subset\kappa}\left(\prod_{v\in\Gamma_\clF} A_v\right)\left(\prod_i A_{\clF_i}\right)\Phi^*_{S_\clF}
\label{OmegaPsigeneralmodelwithtopologicalproperty}
\ee
that satisfies our requirement: each spin foam amplitude is uniquely determined by the branching graph, the coloring and topology of single-colored sheets. Assume also that the boundary functional $\Phi$ has a topological dependence: the coefficient $\Phi_{S_\clF}$ is a function of the connectivity and labelling of the spin network $S_\clF$, and no information on the lattice is needed to compute its value.
In that case, we extend  \eq{OmegaPsigeneralmodelwithtopologicalproperty} to a background independent sum over all spin foams in the manifold:
\be
\Omega(\Phi) = \sum_{\clF\subset M}\left(\prod_{v\in\Gamma_\clF} A_v\right)\left(\prod_i A_{\clF_i}\right)\Phi^*_{S_\clF}
\label{OmegaPsiwithoutlattice}
\ee
(The same idea could be also expressed by a refinement limit: equip the manifold with an auxiliary Euclidean metric, and choose a sequence $\kappa_i$ of triangulations for which the volume of tetrahedra goes uniformly to zero. The formal limit
\be
\lim_{i\to\infty} \Omega_{\kappa_i}(\Phi)
\ee
contains all spin foams on $M$ whose valence does not exceed that of a dual triangulation.)

The transition from \eq{OmegaPsigeneralmodelwithtopologicalproperty} to \eq{OmegaPsiwithoutlattice} requires a generalization of the Hilbert space, since the spin networks $S_\clF$ are no longer restricted to the boundary lattice $\pa\kappa$. We define the new space of boundary states as follows: we take the space
\be
\clH_{\pa M} := \left\{\,\sum_{i=1}^n a_i S_i\;\Big|\; a_i\in\bC\,,\;\;S_i\subset M\,,\;\;n\in\bN\,\right\}
\ee
of finite linear combinations of spin networks in the boundary $\pa M$, equip it with the inner product
\be
\b S, S' \k = \delta_{S,S'}\,.
\ee
and define the space of boundary states as $\clH^*_{\pa M}$, i.e.\ as the space of linear functionals $\Phi: \clH_{\pa M}\to\bC$.

\psfrag{hstarF}{$h^* F$}
\psfrag{arrowh}{$\stackrel{h}{\longrightarrow}$}
\pic{actionofhomeomorphismonspinfoam}{Action of a homeomorphism $h$ on a spin foam $F$.}{3cm}{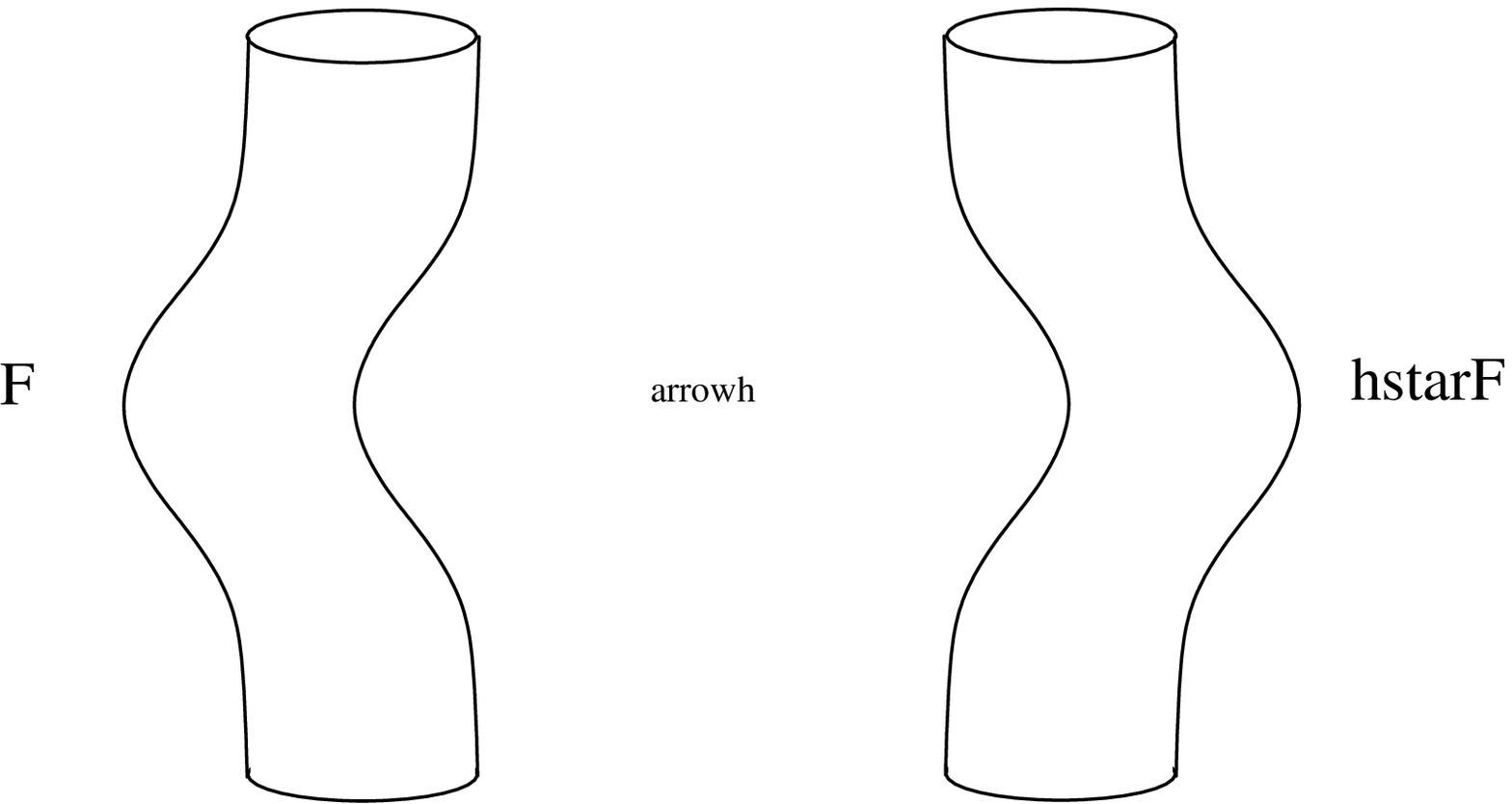}
Clearly, the spin foam sum \eq{OmegaPsiwithoutlattice} contains a huge overcounting, which is due to the fact that amplitudes do not depend on how spin foams are embedded in the manifold (\fig{actionofhomeomorphismonspinfoam}). More precisely, each spin foam amplitude
\be
A(\clF) := \left(\prod_{v\in\Gamma_\clF} A_v\right)\left(\prod_i A_{\clF_i}\right)
\ee
and boundary coefficient $\Phi_{S_\clF}$ is invariant under the action of homeomorphisms $h: M\to M$, which we write as
\be
A(h^*\clF) = A(\clF)\qquad\mbox{and}\qquad \Phi_{S_{h^*\clF}} = \Phi_{S_F}\,.
\ee
To eliminate this overcounting, we factor off infinite gauge volumes à la Fadeev-Popov, and replace \eq{OmegaPsiwithoutlattice} by
a sum over equivalences classes $\Ft$ of spin foams under homeomorphisms: \setlength{\jot}{0.2cm}
\bea
\Omega(\Phi) &=& \sum_{\Ft} \,V({\rm Homeo}(M))\,\At(\Ft)\,\Phit^*_{\St_{\Ft}} \\
\downarrow\hspace{0.4cm} & & \nonumber \\
\Omegat(\Phit) &=& \sum_{\Ft} \At(\Ft)\,\Phit^*_{\St_{\Ft}}\,.
\label{abstractspinfoamsum} 
\eea
We call such equivalence classes {\it abstract} or {\it topological} spin foams. Correspondingly, we define an {\it abstract spin network} as an equivalence class of spin networks under homeomorphisms of the boundary. $\St_{\Ft}$ stands for the equivalence class of spin networks that is induced by $\Ft$. The functionals $\At$ and $\Phit$ are defined by
\be
\left.
\parbox{3cm}{
$\renewcommand{\arraystretch}{1.2}\begin{array}{l}
\At(\Ft) = A(\clF) \\
\Phit(\St_{\Ft}) = \Phi_{S_\clF}
\end{array}\renewcommand{\arraystretch}{1}$
}
\right\}\;\;\mbox{for any representant $\clF$ of $\Ft$.}
\ee
In a more explicit form, equation \eq{abstractspinfoamsum} reads
\be
\Omegat(\Phit) = \sum_{\Ft}\left(\prod_{v\in\Gammat_{\Ft}} A_v\right)\left(\prod_i A_{\Ft_i}\right)\Phi^*_{\St_{\Ft}}
\label{abstractspinfoamsummoreexplicit}
\ee
where each tilded quantity is an equivalence class under homeomorphisms.

The definition of abstract spin foams and spin networks parallels that of abstract spin networks in canonical loop quantum gravity\footnote{See e.g.\ sec.\ 6.4 in \cite{Rovellibook}}, and that of 3- and 4-geometries in the sum-over-metrics approach to gravity. In our case, the dynamics is insensitive to moduli of spin networks, so we extended the symmetry group from diffeommorphisms to homeomorphisms. 
As defined above, abstract spin foams are closely related to the spin foams of group field theories, which are labelled non-embedded 2-complexes. There is a crucial difference, however: the latter can have trivial labels, while, by construction, our spin foams carry only non-trivial representations. We discuss the consequences of that difference in the final section. 
\setlength{\jot}{0cm}

The upshot of all this is the following: by going through the steps from \eq{OmegaPsigeneralmodelwithtopologicalproperty} to \eq{abstractspinfoamsummoreexplicit}, we can start from any spin foam model on a lattice whose amplitudes satisfy topological invariance, and construct a manifestly background free theory from it. The latter is specified by sums over abstract spin foams, which carry only topological and combinatorial information. There is, in particular, a version of the Barrett-Crane model which meets our requirements (see \eq{OmegaPsiBarrettCrane}), so we can extend it to an abstract sum
\be
\Omegat(\Phit) = \sum_{\FBCt}\left(\prod_{v\in\Gammat_{\FBCt}}
\parbox{3cm}{\includegraphics[height=2cm]{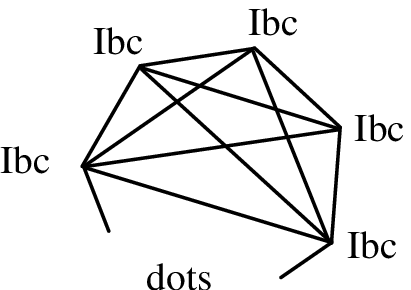}}
\right)\left(\prod_i\left(\dim V_{\rho_i}\right)^{\tilde{\chi}(\FBCt{}_i)}\right)\Phi^*_{\St_{\FBCt}}\,.
\label{OmegaPsiBarrettCraneabstract}
\ee

As such, a sum of the type \eq{OmegaPsiBarrettCraneabstract} is highly divergent: for any given abstract spin network in the boundary, there appears an infinite number of disconnected spin foams, an infinity of topologies for each single-colored sheet, and also infinitely many connected spin foams, due to the presence of bubbles \cite{PerezRovellibubbles}. To arrive at a concrete model, an appropriate truncation or dampening will be necessary: for example, a restriction to connected spin foams, a cutoff on topologies and an exclusion of bubbles.

\section{Summary and discussion}
\label{summaryanddiscussion}

Let us summarize the contents and results of the paper:

We have given a pedagogic derivation of the transformation from pure lattice gauge theory to its dual spin foam model. In doing so, we emphasized the grouping of plaquettes into single-colored surfaces, and were naturally led to a geometric definition of spin foams: spin foams are not identified with the lattice and its labellings, but instead regarded as geometric objects---branched surfaces---that are placed on the lattice. This geometric viewpoint enabled us to write down a very simple spin foam model of lattice Yang-Mills theory for gauge group SU(N) and dimension $d\ge 2$. Its spin foams are weighted with an ``action'' that is proportional to the area of surfaces, similar as worldsheets of the Nambu-Goto string. The proportionality constant depends on the representation label of unbranched sheets and can be viewed as a tension of spin network edges. The running of the original gauge coupling is mapped into a running
$$
T_\rho(a) := a^{(d-6)}\gamma^2(a)\,C_\rho
$$
of the edge tension. It should be stressed that the transformation from gauge theory to spin foam model is non-perturbative, so it does not require a strong coupling expansion.

In the second part of the article (sec.\ \ref{backgroundindependentspinfoammodels}) we applied the notion of geometric spin foams to models of gravity: we introduce a symmetry condition that requires spin foam amplitudes to be independent of the lattice. There are two motivations for this step: firstly, some of the existing models meet the symmetry requirement, or do so after a simple modification of the amplitude. Secondly, it allows for a purely geometric definition of the spin foam sum: for each spin foam, the amplitude depends only on the topology and labelling of the branched surface, and the lattice cutoff translates into the condition that we should only sum over spin foams that lie on the lattice. In models that have the desired symmetry property, we discard the lattice regularization and extend the sum to all spin foams in the manifold. After factoring off infinite gauge volumes (the volume of the homeomorphism group), we arrive at a sum over abstract spin foams. Thus, we obtain a purely combinatorial model that is free of any choice of lattice or triangulation. Our procedure applies to a version of the Barrett-Crane model, as we show in the appendix.

The two parts of the article are tied together by the idea that spin foams should be regarded as geometric objects. Strictly speaking, however, the gauge and gravity case are treated on a different footing: in Yang-Mills theory, we picked out a dual model that has a nice geometric interpretation, but that does not prevent us from using other non-geometric models (e.g.\ the model associated to the Wilson action) to describe the same low-energy physics. For the gravity models, on the other hand, we impose the symmetry condition from the start, and exclude models that do not satisfy it. We do that, because a lattice dependence of amplitudes would go against the idea of background independence, so we need amplitudes that are only determined by geometric properties of the spin foam.

\subsubsection*{Yang-Mills spin foam model}

The construction of the Yang-Mills model rests on two inputs: the simplicity of the character coefficients of the heat kernel action, and the combination of plaquettes into larger single-colored surfaces. Although both of this is known in lattice field theory and quantum gravity, we have not found any previous definition of the model in the literature. It would be interesting to include fermions and see how they add into the geometric picture of the model. Could it provide an alternative method in the non-perturbative analysis of QCD? A great deal of research has been devoted to the relation between Yang-Mills theories and string theory: there is the ``old'' idea of describing lattice gauge theory in terms of an effective string theory\footnote{For references, see e.g.\ \cite{Muensterstrings}.}, and the more recent program on the correspondence between supersymmetric Yang-Mills theory and superstring theory \cite{Maldacena}. As we have seen, our dual model shares features with the Nambu-Goto string, so can it tell us anything about the string-gauge theory relation? What is its large $N$ limit?

Another interesting question is if the model could be combined with gravity spin foams to give a coupling of Yang-Mills theory and gravity. That possibility is suggested by the area weighting of the Yang-Mills spin foams and the interpretation of gravity spin foams as quanta of area. 

\subsubsection*{Geometric spin foams versus group field theory}

In the context of gravity, our geometric viewpoint suggests a natural way to overcome the triangulation dependence of spin foam models. We consider only models whose weights are functions of topological properties of the spin foam: i.e.\ we have a rule that determines the amplitude for a given branching graph, coloring and topology of surfaces, without making reference to the underlying triangulation. When a model has this property, we keep the 
rule for the amplitudes, but replace the sum over spin foams on the triangulation by a sum over abstract spin foams---topological equivalence classes of spin foams. The resulting new model is manifestly background-independent and contains no information on the triangulation we started from. 

Thus, our approach provides an alternative to the group field theory method, where the triangulation dependence is removed by summing over a large class of complexes. The sum over complexes results from a sum over Feynman diagrams in the perturbative expansion of a group field theory (GFT). The graph of a Feynman diagram corresponds to an abstract 2-complex, and its value is a spin foam sum on the complex, i.e.\ a sum over irrep labels on faces and intertwiners on edges. In this context, a spin foam is identified with the labelled, abstract complex, so we can view the entire expansion as a sum over abstract spin foams. 

What is the relation between this sum and the sum over abstract spin foams we have defined? The answer is that they are very different, and the reason for it lies in different definitions of spin foams: in the GFT sense, a spin foam is an abstract, non-embedded 2-complex, together with irrep labels on faces and intertwiner labels on edges. Trivial irreps are {\it allowed}.

In our approach, we also arrived at abstract spin foams, but we started from the geometric concept of spin foam, where faces of the same color are merged into larger surfaces, and trivial labels are ignored. Therefore, our abstract spin foams are equivalent to non-embedded 2-complexes where trivial irreps are {\it excluded}.

This difference has dramatic consequences for the spin foam sum: in the GFT expansion, one sums over all labelled complexes, no matter how much or little of the complex is labelled trivially. In particular, the same non-trivial part can appear in infinitely many spin foams which just result from adding trivial parts to it. The weights of these spin foams differ only by symmetry factors and powers of the coupling constant, while the actual spin foam amplitude stays the same. In our approch, the non-trivial part is only counted once, since trivial labels do not appear in the bookkeeping.

For that reason, the present proposal for abstract spin foam sums seems better suited for an interpretation in terms of histories of spin networks: here, a history of non-trivial spin networks is counted once, while the GFT counts also all ways of adding trivial spin networks to it.

\subsubsection*{Semiclassical analysis}

A central problem in the research on spin foam models is the semiclassical limit. We would like to know if theories like the Barrett-Crane model can produce any physically realistic low-energy limit. Can it generate backgrounds that resemble gravity? Does it exhibit critical behaviour? Following the ideas in \cite{Smolinstringperturbations}, we can sketch a possible way to tackle these questions: we view abstract spin foams as classical configurations, and the associated amplitudes as a kind of exponentiated action $\exp\left(\irm\,\clS\right)$. In the space of abstract spin foams, we have a certain notion of continuity, which is given by incremental relabelling and stepwise modification of spin foam topologies. Thus, we can define variations of the action with respect to spin foam configurations. The analogy with field theory suggests that backgrounds might be identified as large spin foam complexes that are extrema of the action. If such extrema exist, one could try to formulate perturbative expansions around them.

\subsection*{Acknowledgements}
I thank Alejandro Perez, Carlo Rovelli and Lee Smolin for comments and discussions.
I also thank the members of the CPT, Marseille for their hospitality.
This work was supported by the Daimler-Benz foundation and DAAD.

\begin{appendix}

\section{Topological invariance of Barrett-Crane amplitudes}

In the original paper on their model, Barrett and Crane fixed only the vertex amplitude, while the amplitudes for edges and faces remained unspecified. Since then various versions of the Barrett-Crane model have been proposed that differ in the choice of edge and face amplitudes. We show below that the model B in \cite{DePietriFreidelKrasnovRovelli} is the version which satisfies our requirement of topological invariance.

We start from the conventional formulation where spin foams are viewed as oriented 2-complexes $\kappa$ with intertwiner labels on edges and irrep labels on faces. Representations are allowed to be trivial.
The choice of $\kappa$ is fixed to a single complex---the dual 2-skeleton of a triangulation.
Then, the partition function is given by
\be
Z = \left(\prod_{f\in\kappa}\,\sum_{\rho^{\sf bal}_f}\right)\left(\prod_{f\in\kappa}\dim V_{\rho^{\sf bal}_f}\right)\left(\prod_{v\in\kappa} A_v\right)
\label{modelB}
\ee
where the sum runs over all possible ways to assign balanced representations of SO(4) to faces. The vertex amplitude results from a contraction of normalized Barrett-Crane intertwiners:
\be
A_v = \parbox{2.6cm}{\includegraphics[height=2cm]{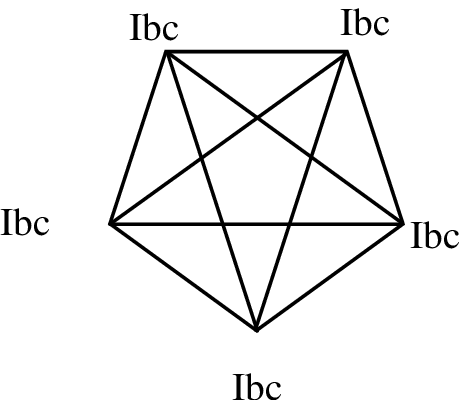}}
\label{vertexdiagram}
\ee
Each vertex of the pentagon corresponds to an edge of the complex, and each edge carries a balanced representation from a face of the complex.

Let us translate this to the geometric language with unbranched components and branching lines.
Within an unbranched component, faces are non-trivially labelled and arranged in a surface. Faces that ``go off'' have the trivial label. In the vertex diagram \eq{vertexdiagram}, this corresponds to having an open or closed line of non-trivial irreps, while other edges are trivial (see \fig{vertexamplitudeinsinglecoloredcomponent}).
\pic{vertexamplitudeinsinglecoloredcomponent}{Vertex amplitude in an unbranched component.}{3cm}{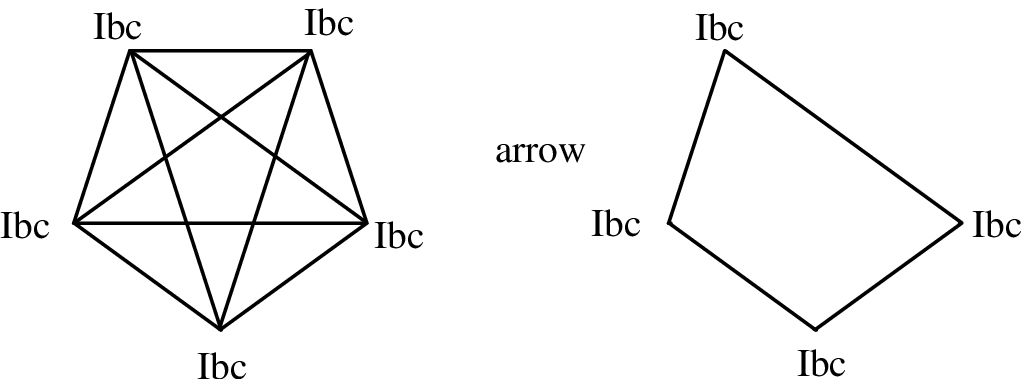}

\psfrag{j1}{$j_1$}
\psfrag{j2}{$j_2$}
\psfrag{djj}{$\delta_{j_1,j_2}$}
To evalute the diagram, we use the definition of the Barrett-Crane intertwiners (see e.g.\ \cite{PfeifferconnectionpictureBarrettCrane}). Recall that a balanced representation of SO(4) is isomorphic to a tensor product $j\otimes j^*$ where $j$ is a representation of SU(2).
A BC-intertwiner
between balanced irreps can be expressed as a Haar intertwiner of the associated SU(2) irreps.
In the 2-valent case, we get
\be
\parbox{14cm}{\includegraphics[height=3cm]{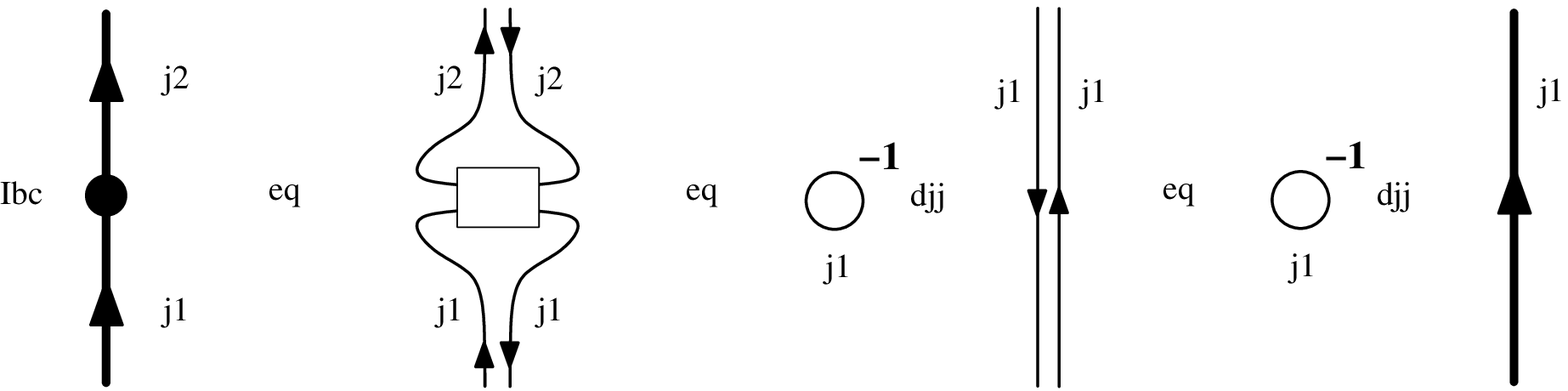}}
\label{twovalentBCintertwiner}
\ee
The thick lines represent contractions w.r.t.\ balanced representations, while thin lines stand for the SU(2) irreps.
We see that \eq{twovalentBCintertwiner} is indeed a {\it normalized} intertwiner, since
\psfrag{j}{$j$}
\psfrag{1}{1}
\psfrag{Ibcstar}{$I^{\ssst\rm BC}{}^*$}
\be
\parbox{7cm}{\includegraphics[height=3cm]{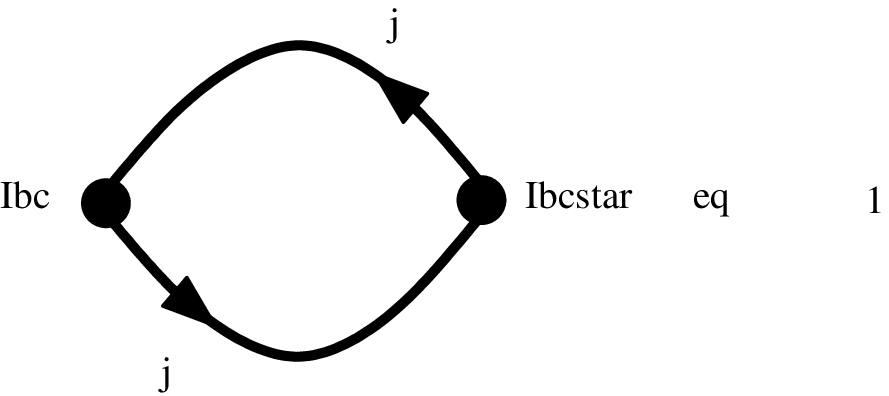}}
\ee
The identity \eq{twovalentBCintertwiner} tells us that in \fig{vertexamplitudeinsinglecoloredcomponent} all representations along a closed line have to be the same, and that open lines would give zero. As in section \ref{dualtransformationtothespinfoammodel}, we conclude that unbranched components have to be single-colored and cannot have open ends. For a closed line, we evaluate the contraction of 2-valent BC-intertwiners and obtain $(\dim V_j)^{-1}$ for each intertwiner and $(\dim V_j)^{+2}$ from the contraction.
The intertwiner on a lattic edge appears in two vertex amplitudes, so an edge contributes a factor
\be
(\dim V_j)^{-2}\,.
\ee
The contraction gives
\be
(\dim V_j)^{+2}
\ee
for each vertex, and the face factor in \eq{modelB} yields the dimension of the balanced representation, which is
\be
(\dim V_j)^{+2}\,.
\ee
For an unbranched component $\clF_i$ with color $\rho^{\sf bal}_i$, the factors add up to
\be
\left(\dim V_{\rho^{\sf bal}_i}\right)^{\chi(\clF_i)}\,.
\ee
It remains to check that contributions from the branching graph are topological. Vertices on the branching line contribute a sequence of factors
\be
\parbox{7cm}{\includegraphics[height=2cm]{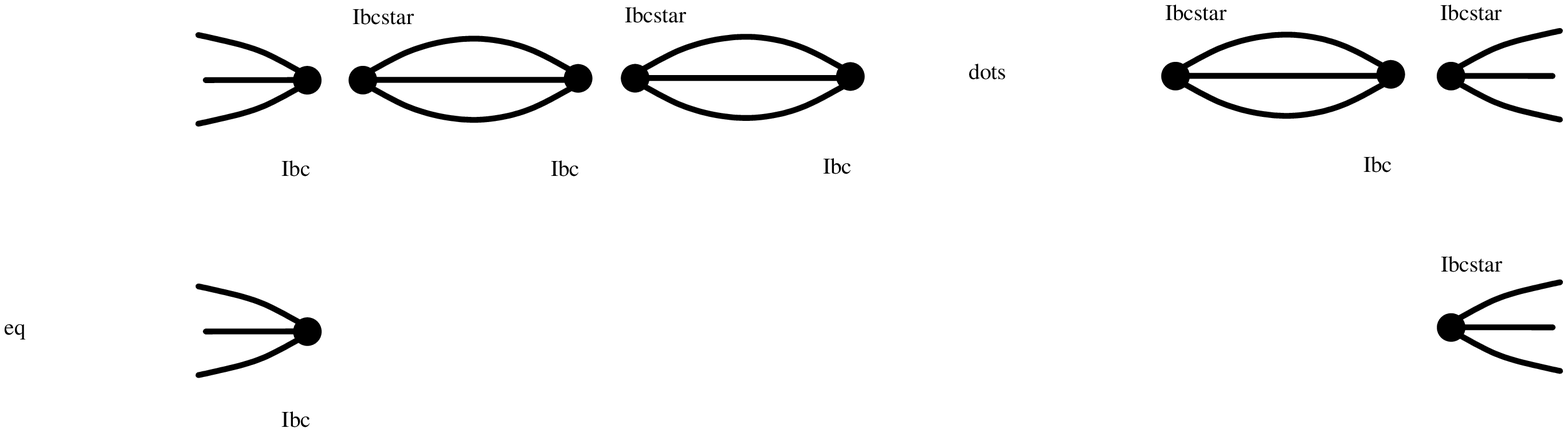}}
\ee
Since the BC-intertwiners of the model are normalized, the sequence collapses to a single pair of BC-intertwiners,
each of which is associated to a vertex of the branching graph. Thus, we arrive at the geometric form of the spin foam sum that was already stated in equation \eq{OmegaPsiBarrettCrane}.

\end{appendix}

\end{document}